\documentclass[prd,aps,a4paper,superscriptaddress,twocolumn,nofootinbib]{revtex4} %

\usepackage{graphicx}
\usepackage{color}
\usepackage{dcolumn}
\usepackage{bm}
\usepackage{slashed}
\usepackage{amsmath}
\usepackage{latexsym}
\usepackage{amssymb}
\usepackage{mathrsfs}
\usepackage{amsfonts}
\usepackage{url}
\allowdisplaybreaks
\graphicspath{{fig/}}
\usepackage{diagbox}
\usepackage{multirow}
\usepackage{makecell}
\usepackage{natbib}

\begin{document}
	\title{Matched filtering for gravitational wave detection without template bank driven\\by deep learning template prediction model bank}
	
	\author{CunLiang Ma}
	\affiliation{School of Information Engineering, Jiangxi University of Science and Technology, Ganzhou, 341000, China}
	
	\author{Sen Wang}
	\affiliation{School of Information Engineering, Jiangxi University of Science and Technology, Ganzhou, 341000, China}
	
	\author{Wei Wang}
	\affiliation{School of Computer Science, Fudan University, Shanghai 201203, China}
	
	\author{Zhoujian Cao
		\footnote{corresponding author}} \email[Zhoujian Cao: ]{zjcao@amt.ac.cn}
	\affiliation{Institute for Frontiers in Astronomy and Astrophysics, Beijing Normal University, Beijing 102206, China}
	\affiliation{Department of Astronomy, Beijing Normal University, Beijing 100875, China}
	\affiliation{School of Fundamental Physics and Mathematical Sciences, Hangzhou Institute for Advanced Study, UCAS, Hangzhou 310024, China}
	
	\begin{abstract}
		The existing matched filtering method for gravitational wave (GW) search relies on a template bank. The computational efficiency of this method scales with the size of the templates within the bank. Higher-order modes and eccentricity will play an important role when third-generation detectors operate in the future. In this case, traditional GW search methods will hit computational limits. To speed up the computational efficiency of GW search, we propose the utilization of a deep learning (DL) model bank as a substitute for the template bank. This model bank predicts the latent templates embedded in the strain data. Combining an envelope extraction network and an astrophysical origin discrimination network, we realize a novel GW search framework. The framework can predict the GW signal’s matched filtering signal-to-noise ratio (SNR). Unlike the end-to-end DL-based GW search method, our statistical SNR holds greater physical interpretability than the $p_{score}$ metric. Moreover, the intermediate results generated by our approach, including the predicted template, offer valuable assistance in subsequent GW data processing tasks such as parameter estimation and source localization. Compared to the traditional matched filtering method, the proposed method can realize real-time analysis. The minor improvements in the future, the proposed method may expand to other scopes of GW search, such as GW emitted by the supernova explosion.
	\end{abstract}
	
	\maketitle
	
	\section{Introduction}
	The direct observation of gravitational waves (GWs) holds immense value for both validating and refining physical theories, as well as advancing our understanding of cosmology. Managed by LIGO-VIRGO-KAGRA (LVK) collaboration, the second-generation ground-based GW detectors have successfully identified more than 90 confident events originating from compact binary coalescences (CBCs) throughout their first, second, and third observing runs \cite{1,2,3,4}. These data have effectively substantiated general relativity \cite{5} and facilitated the measurement of the Hubble constant \cite{6}. Additionally, these data have been employed to explore evidence of cosmic strings \cite{7}, continuous GWs \cite{8}, and stochastic GW backgrounds \cite{9}. For differences in the detectable frequency bands, space-based GW detectors \cite{10, 11, 12} and third-generation (3G) ground-based detectors \cite{13,14} will substantially enhance the diversity and quantity of astrophysical sources associated with GW. Consequently, the GW search pipeline must continually improve to meet the forthcoming challenges.
	
	The LVK organization employs PyCBC \cite{15}, GstLAL \cite{16}, MBTA \cite{17}, SPIIR \cite{18}, and CWB \cite{19} for the searches of GW signals. The CWB uses a non-template prior method, whereas the other four pipelines are all based on the template based matched filtering (MF) technique. Although MF offers the benefit of being physically interpretable, it suffers from the drawback of having low computational efficiency due to the large size of the template bank.  In the case of the higher-order modes and eccentricity playing an important role in the 3G era, current template-bank-based algorithms will hit computational limits. 
	
	To speed up the MF-based GW search method, in this work, a deep learning (DL) model bank is proposed to substitute the template bank. The templates are predicted by the DL model bank rather than contained in a template bank. Notably, the outputs of the proposed method are also matched filtering signal-to-noise ratio, and the advantage of the proposed method is the accelerated computational speed.
	
	DL methods for GW detection have been extensively investigated. Back in 2018, George et al. \cite{20} and Gabbard et al. \cite{21} independently showcased the potential of a DL-based algorithm for GW detection. Thus far, the DL-based approaches for GW detection have undergone significant development \cite{22,23,24,25,26,27,28,29,30,31,32,33,34}. Furthermore, there also have been various endeavors attempts to utilize deep learning to other aspects of GW data processing, such as localizing the astrophysical origin \cite{35,36} and estimating the parameters of GW sources \cite{37,38,39,40}. 
	
	Like computer vision \cite{41} and nature language processing \cite{42} tasks, DL-based GW detection can often exhibit a black-box nature. This characteristic makes the end-to-end DL models for GW detection less suitable for making statistically significant claims about gravitational wave detections \cite{43}. One contribution of our work is that we propose a novel scheme for GW search with feedback.  Instead of directly performing the end-to-end GW detection task, we use DL models for template prediction, and these predicted templates are then integrated into the conventional MF (Matched Filtering) based method.
	
	The results show that most of the confident events in GWOSC-1, GWOSC-2, GWOSC-2.1, and GWOSC-3 can be identified. Notably, the proposed method is physically interpretable because other than the $p_{score}$ in the previous DL-based detection method \cite{20,21}, the event significance is measured by the signal-to-noise ratio and the significance of each coincident trigger can be estimated by calculating the false alarm rate (FAR).
	
	Most of the time slices only contain background noise. Almost all the DL-based GW detection methods have a disadvantage in using a time-sliding method with a small time sliding step \cite{28}. To address this drawback, we use the coalescence time information predicted by the envelope extraction model \cite{31} to align the data analysis window. When the coalescence time is obtained, the potential templates will be predicted by the models in the denoising model bank. 
	
	The potential templates derived from the denoising model cannot be directly employed for MF, due to the possibility that the denoised output might not accurately resemble the binary black hole (BBH) GW shape. To address this concern, we propose the astrophysical origin identification models to determine whether the denoised waveform corresponds to the BBH GW shape.  The potential templates are selected by these proposed identification models. 
	
	The temporal extent of the GW signal within the sensitivity band of the LIGO-VIRGO-KAGRA network is influenced by the masses of the binary system. For BBH systems, this duration ranges from a fraction of a second for higher masses to a few seconds for lower masses. Recognizing this variability, we organize the denoising model bank to account for specific sub-parameter spaces. These sub-spaces are defined by dividing the parameter range based on the binary system's masses. Consequently, each model in the denoising model bank is focused on templates within a particular sub-parameter space. The input duration for denoising models is then adjusted according to the corresponding sub-parameter space.
	
	This paper is organized as follows. The proposed framework and the stages in the framework are detailed in Section II. In Section III, we describe the training datasets and training schemes for the neural networks in the proposed framework. In Section IV, we describe the results of the stages tested by the test dataset, the real confident events, and half a month of detected data in Hanford. The conclusion and discussion are detailed in Section V.
	
	\section{\label{sec2}THE PROPOSED FRAMEWORK FOR GW SEARCH}
\begin{figure*}
		\includegraphics[scale=0.54]{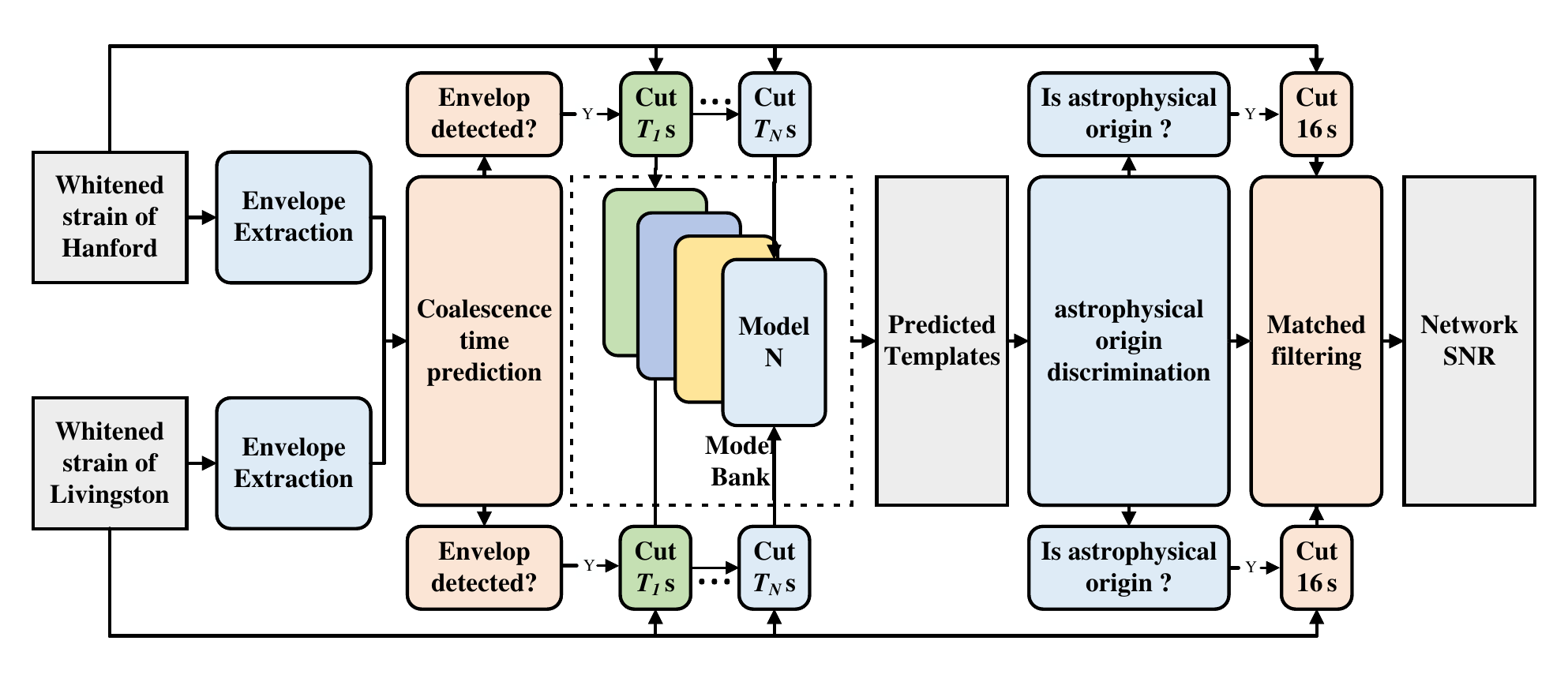}
		\caption{\label{fig:framework}The overview of the proposed GW detection framework.}
	\end{figure*}

	Our fundamental purpose is to construct an interpretable framework for GW search. To alleviate the computational overhead issue of the MF-based method, we propose a muti-step processing method, which is shown in Fig.~\ref{fig:framework}. The whole framework consists of four stages, significant time prediction (STP), the preliminary templates prediction (PTP), template selection (TS), and matched filtering (MF). The details of the four stages will be introduced in the following four sub-sections.
	
	\subsection{\label{sec:level2}The significant time prediction stage}
	
	The output of the STP stage is a time set $S_t=\left\{t_1, t_2, t_3,\dots \right\},t_i$ is the significant time. These significant times correspond to the coalescence time predicted by the envelope prediction network.  As most of the strain data only contain background noise, most of the time, $S_t=\varnothing$. We feed the whitened strain $\boldsymbol{s}_X$(the subscript $X\in{H,L}$ denoting the Hanford or Livingston interferometer) to the envelope extraction network. The relation between $\boldsymbol{s}_X$ and the output of the envelope prediction network ($\hat{\boldsymbol{h}}_X^{env}$) can be denoted as
	\begin{equation}
		\hat{\boldsymbol{h}}_X^{env}=EnvNet(\boldsymbol{s}_X \mid \boldsymbol{W}_{env}),\label{eq1}
	\end{equation}
	where $EnvNet$ represents a parameterized system which is proposed in \cite{31}, and $\boldsymbol{W}_{env}$ represents the envelope extraction model’s trainable weights. If max$(\hat{\boldsymbol{h}}_X^{env})>0.5$, then the coalescence time of interferometer $X$ can be predicted by
	\begin{equation}
		t_X = \Delta t \times \mathop{\arg\max}\limits_{n} \hat{h}_X^{env}(n),\label{eq2}
	\end{equation}
	where $\Delta t$ is the sampling period, and $\hat{h}_X^{env}(n)$ represents the predicted envelope amplitude at time $n\times \Delta t$. In this work, we employ the envelope extraction model trained in \cite{31} for direct envelope prediction. The envelope extraction method proposed in \cite{31} is regarded to test the coalescence time of different detectors (Hanford and Livingston). The difference is that, in this work, the network is used to quickly focus on the potential time segments that may contain GWs. If $|t_H-t_L |<0.05\ s$, then $t^s=\frac{t_H+t_L}{2}\in S_t$. Otherwise, if $t_X$ exists, then  $t_X \in S_t$.
	
	\subsection{\label{sec:level2}The preliminary templates prediction stage}
	
	In the case of $S_t\neq \varnothing$, the PTP stage will work. The outputs of the PTP stage are two templates set $S_{PT}^{H}=\{\hat{\boldsymbol{h}}_H^1,\hat{\boldsymbol{h}}_H^2,\hat{\boldsymbol{h}}_H^3,\dots\}$ and $S_{PT}^{L}=\{\hat{\boldsymbol{h}}_L^1,\hat{\boldsymbol{h}}_L^2,\hat{\boldsymbol{h}}_L^3,\dots\}$, where $\hat{\boldsymbol{h}}_X^i$ denotes the \emph{i}-th preliminary predicted templates. In the PTP stage, a DL model bank is used. The DL model bank is a set of DL models for denoising. The DL model bank can be denoted as
	\begin{equation}
		S_{DM} = \left\{DenoiseModel_i\mid i\in [1,N_M]\right\},\label{eq3}
	\end{equation}
	where $N_M$ is the number of the denoising models. Suppose the input of $DenoiseModel_i$ is $\boldsymbol{s}_X^i\in \mathbb{R}^{1\times M_i}$, where $M_i=T_i\times 4096$, $T_i$ is the time duration of the input strain of $DenoiseMode_i$. The output can be denoted as
	\begin{equation}
		\hat{\boldsymbol{h}}_X^i = DenoiseModel_i\left(\boldsymbol{s}_X^i\mid \boldsymbol{W}_{Mod_i}\right),\label{eq4}
	\end{equation}
	$\hat{\boldsymbol{h}}_X^i$ is the denoised strain. $\boldsymbol{W}_{Mod_i}$ denotes the $DenoiseModel_i$'s trainable weights, which can be optimized by
	\begin{eqnarray}
		\boldsymbol{W}_{Mod_i}=&&\mathop{\arg\min}\limits_{\boldsymbol{W}_{Mod_i}}\frac{1}{N_T}\sum\|DenoiseModel_i\nonumber\\&&\left(\boldsymbol{s}_X^{i,k}\mid \boldsymbol{W}_{Mod_i}\right)-\boldsymbol{h}_X^{i,k}\|^2,\label{eq5}
	\end{eqnarray}
	where $\boldsymbol{h}_X^{i,k}$ denotes the whitened signal buried in the \emph{k}-th whitened strain sample. 
	
	In contrast to recent approaches in GW denoising, like WaveNet \cite{44}, LSTM \cite{45}, and WaveFormer \cite{46}, our methodology employs a multitude of models. Each denoising model only focuses on a subset of the parameter space. The source parameter of the BBH is divided by the mass range of the binaries. As we know, the timescale of the GW signals within the sensitivity band of the detector network varies according to the BBH masses of the system. For the PyCBC pipeline, the candidate and background events are divided into three search classes based on template length \cite{47,48}. Motivated by this, the input length of the denoising model varies with the concerned parameter range. The sub-parameter spaces and the denoising models’ input duration are listed in Table~\ref{tab:table1}.
	\begin{table}[htb]
		\caption{\label{tab:table1}The sub-parameters and input time length of denoising models. The unit of the BBH’s mass is $M_\odot$.}
		\begin{center}
			\begin{ruledtabular}
				\begin{tabular}{ccccc}
					\diagbox{$m_2$}{$m_1$}&(5,10]&(10,20]&(20,40]&(40,80]\\
					\hline
					(5,10]                &2.00 s&1.75 s &1.50 s &1.25 s \\
					(10,20]               &\dots &1.50 s &1.00 s &0.75 s \\
					(20,40]               &\dots &\dots  &0.75 s &0.50 s \\
					(40,80]               &\dots &\dots  &\dots  &0.25 s \\
				\end{tabular}
			\end{ruledtabular}
		\end{center}
	\end{table}
	
	Since the effectiveness of the U-Net like model for the envelope extraction task in \cite{31} we also adopt the U-Net like model for denoising. The input shape of the denoising models varies with the parameter space of the astrophysical origin of the GW signal. The model structure of all the denoising models in the model bank is the same. There are two differences between different models. The differences are the parameter space of training data and the input shape.
	\begin{figure*}
		\includegraphics[scale=0.645]{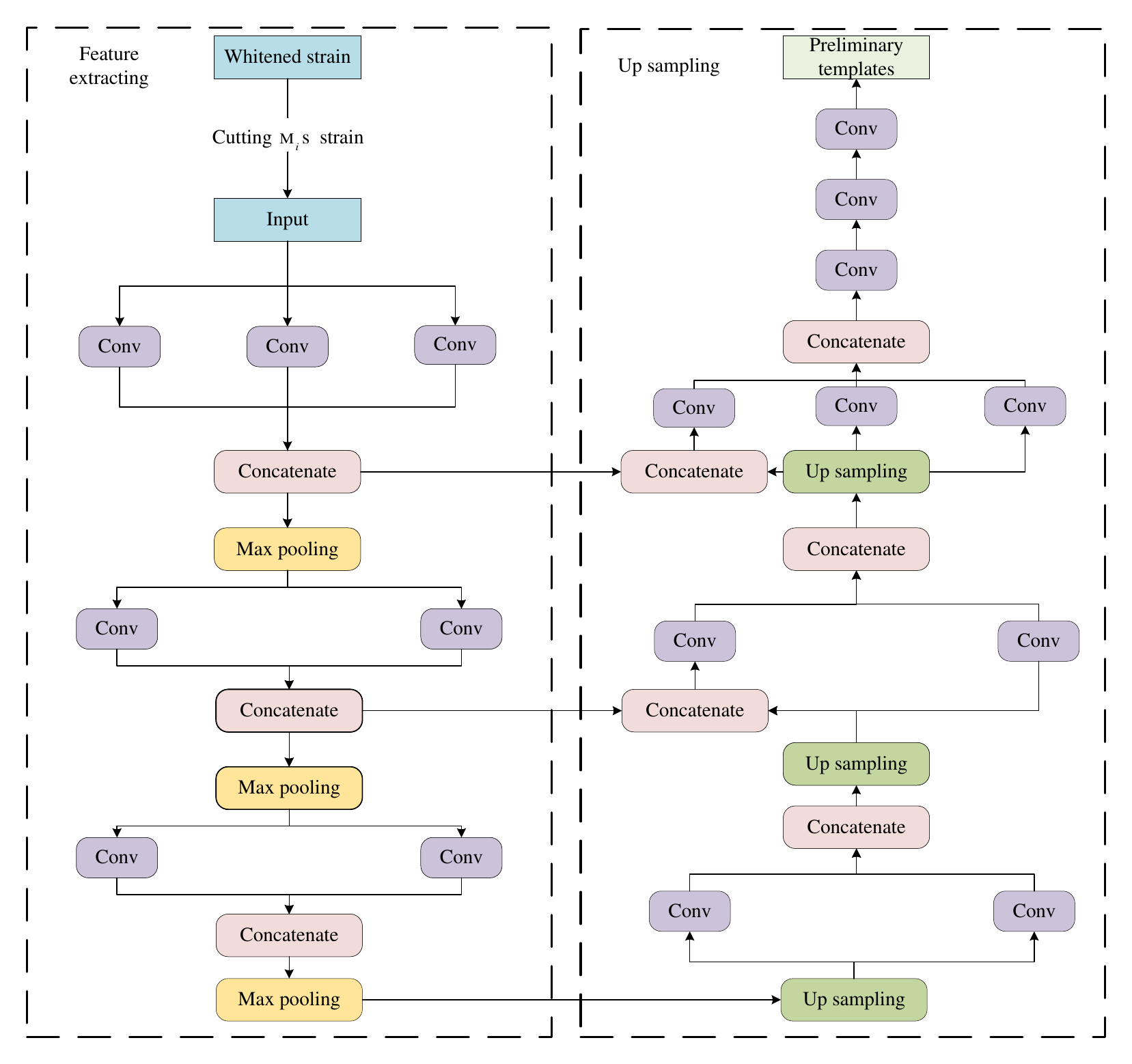}
		\caption{\label{fig:u_net}Network structure of the denoising models.}
	\end{figure*}
	
	\subsection{\label{sec:level2}The template selection stage}
	
	The astrophysical origin discrimination networks are used in this stage. The inputs of this stage are the preliminary templates $S_{PTP}^X=\{\hat{\boldsymbol{h}}_X^1,\dots,\hat{\boldsymbol{h}}_X^{N_M}\}$, and the outputs of this stage are the selected templates that will be used in the matched filtering stage. Suppose the selected templates comprise a set $S_{Tem}^X$. If a preliminary template generated by the denoising model does not have a GW waveform shape, the matched filtering results will not reflect the GW’s SNR information. Therefore, before the matched filtering stage, we select the templates with GW waveform shape.
	
	We constructed astronomical origin discrimination networks $\{AstroDisNet_1, AstroDisNet_2,\dots, AstroDisNet_{N_M}\}$ to do this work. An astrophysical origin discrimination network is a classification model which can distinguish between the waveform with an astronomical origin or not.  Take $\hat{\boldsymbol{h}}_X^i$ and plug it into $AstroDisNet_i$, then the relation between the input and output can be denoted as 
	\begin{equation}
		\boldsymbol{o}_{astro}^i=AstroDisNet_i(\hat{\boldsymbol{h}}_X^i\mid \boldsymbol{W}_{astro_i}),\label{eq6}
	\end{equation}
	where $\boldsymbol{W}_{astro}$ is the trainable weights and is optimized by
	\begin{eqnarray}
		\boldsymbol{W}_{astro_i}=&&\mathop{\arg\min}\limits_{\boldsymbol{W}_{astro_i}}\frac{1}{N_{a_i}}\nonumber \\
		&&\sum_{k}CrossEntropy(\boldsymbol{y}^k,\boldsymbol{o}_{astro}^{i,k}),\label{eq7}
	\end{eqnarray}
	when the \emph{k}-th sample has an astrophysical origin shape, $y^k=1$, otherwise, $y^k=0$. In this work, we use a convolutional network to utilize $AstroDisNet_i$. Every network only focuses on the waveform originating on a sub-parameter space. When $\boldsymbol{o}_{astro}^i$ of both Hanford and Livingston are all bigger than a threshold $o_{th}$, the $\hat{\boldsymbol{h}}_X^i$ will be adopted as a template and $\hat{\boldsymbol{h}}_X^i\in S_{Tem}^X$. Note that the number of templates in  $S_{Tem}^H$ and $S_{Tem}^L$ are the same.
	
	For being widely studied in recent years \cite{49,50,28,29}, we adopt the GW detection model designed by Gabbard et al. \cite{21} as the structure of the astrophysical origin discrimination network shown in Table~\ref{tab:table2}. Because the time length of the input varies with the focused parameter space, the input shape of the networks is different. We apply the Elu activation function after every layer except the last dense layer. For each convolutional layer, no padding is applied. A soft-max activation function is used after the last dense layer. 
	\begin{table*}[htbp]
		\centering
		\caption{\label{tab:table2}The structure of the astronomical origin discrimination network.}
		\begin{ruledtabular}
			\begin{tabular}{ccccccc}
				Layer&Layer Type&No.Neurons&Filter size&Max pool size&Dropout&Activation function\\
				\hline
				1    &Convolution&32       &32         &Not applicable&0     &Elu                \\
				2    &Convolution&32       &32         &8             &0     &Elu                \\
				3    &Convolution&64       &16         &Not applicable&0     &Elu                \\
				4    &Convolution&64       &16         &4             &0     &Elu                \\
				5	 &Convolution&128      &8          &Not applicable&0     &Elu                \\
				6    &Convolution&128      &8          &4             &0     &Elu                \\
				7	 &Dense      &64       &Not applicable&Not applicable&0.5&Elu				 \\
				8	 &Dense      &32       &Not applicable&Not applicable&0  &Elu				 \\
				9    &Dense      &2        &Not applicable&Not applicable&0  &Softmax		     \\
			\end{tabular}
		\end{ruledtabular}
	\end{table*}
	
	\subsection{\label{sec:level2}The matched filtering stage}
	
	If the set of the selected template $S_{Tem}\neq \varnothing$, the matched filtering stage will work. The output of this stage is two SNR sets $S_{snr}^H=\{SNR_1^H,SNR_2^H,SNR_3^H,\dots \}$ and $S_{snr}^L=\{SNR_1^L,SNR_2^L,SNR_3^L,\dots \}$. Here we detail the method for obtaining $S_{snr}^X$ (interferometer $X$). For the template $h_i^X (t)\in S_{Tem}$, the output of the matched filtering can be written as
	\begin{equation}
		\rho_{i,X}^2(t)=\frac{\langle d \mid h_i^X\rangle (t)}{\langle \boldsymbol{h}_i^X \mid \boldsymbol{h}_i^X \rangle},\label{eq8}
	\end{equation}
	where $d(t)$ is the strain. $\langle d\mid h_i^X\rangle (t)$ denotes the time-dependent inner product of $d(t)$ and $h_i^X (t)$. In this work, the $\boldsymbol{d}$ and $\boldsymbol{h}_i^X$ are all whitened by the amplitude spectrum density (ASD) of the background noise, so
	\begin{equation}
		\langle d\mid h_i\rangle(t)=4\int_{0}^{\infty}\widetilde{d}(f)\widetilde{h}_i^{X^*}e^{-2\pi jft}df,\label{eq9}
	\end{equation}
	and
	\begin{equation}
		\langle h_i\mid h_i\rangle=4\int_{0}^{\infty}\widetilde{h}_i^X(f)\widetilde{h}_i^{X^*}(f)df,\label{eq10}
	\end{equation}
	where $\widetilde{d}(f)$, $\widetilde{h}_i^X(f)$ are the Fourier transform results of $d(t)$ and $h_i^X (t)$, and $\widetilde{h}_i^{X^*}(f)$ is the conjugate function of $\widetilde{h}_i^X(f)$. Then the network SNR can be calculated by
	\begin{equation}
		NetworkSNR=\mathop{\max}\limits_i \sqrt{\rho_{i,H}^2+\rho_{i,L}^2}.\label{eq11}
	\end{equation}
	
	\section{\label{sec3} TRAINING METHODS}
	
	In the previous section, the proposed framework is detailed. In this section, we will detail the data for training the models and the schemes for training.
	
	\subsection{\label{sec:level2} Data for training the denoising model}
	
	We constructed 10 training data sets (corresponding to 10 sub-parameter spaces shown in Table~\ref{tab:table1}). For each sub-parameter space, 100 thousand samples are generated for training. The background noise’s GPS time was sampled from 1238163456 to 1238806528. Every sample in the training set contains one background noise time series $n(t)$ and one GW signal time series $h(t)$ originating at the BBH system. Both the $n(t)$ and $h(t)$ were whitened by the noise ASD which is estimated by the Weltch method. The time duration of  $n(t)$ and $h(t)$ are all 16 s, and the sampling rate is 4096 Hz.
	
	The waveforms were generated using the IMRPhenomD model for the case of at least one of the black hole’s masses being less than $10\ M_\odot$. Otherwise, we used the SEOBNRv4 model to simulate the GW waveforms. In the waveform simulation, we sampled the dimensionless spin randomly in (0,0.998). The polarization angle and coalescence phase were sampled randomly in the interval $(0,2\pi)$. The declination and the right ascension were sampled uniformly over a sphere, and the inclination angle was sampled from a uniform distribution from the interval $(0, \pi)$.  
	
	\subsection{\label{sec:level2} Training scheme for the denoising model}
	
	The training scheme for the denoising model is improved compared to the previous work \cite{31}. In \cite{31}, every data sample’s SNR was fixed, because the noise and the signal were added before training. In this work, for each batch, the noise and the signal were synthesized at the training stage as
	\begin{equation}
		\boldsymbol{s}=\lambda \times h(t_b^h:t_e^h)+n(t_b^n:t_e^n),\label{eq12}
	\end{equation}
	where $\lambda$, $t_b^h$, and $t_b^n$ are the three data augmentation factors. $\lambda$ can control the SNR of the trained sample. $t_b^h$ and $t_b^n$ can control the selected time slices for training.  $t_e^h-t_b^h$ and $t_e^n-t_b^n$ are the same value and are equal to the input time duration of the trained model for one generated sample. The time slice of background noise was sampled randomly from the whole 16 s length data for every batch of training. Each time slice of the pure signal is randomly cut, and the coalescence time is within 60\% to 95\% of the time window.  In the training stage, the  $\lambda$ is changed to fit an SNR range for every sample. The SNR ranges change in sequence according to [(17, 20), (14, 17), (10, 14), (5, 10), (5, 10), (10, 14), (14, 17)]. The Adam algorithm is applied to optimize the model parameters, the learning rate is set to 1.5e-5, and the batch number value is 64, for a total of 30 epochs trained.
	
	\subsection{\label{sec:level2} Data for training the astrophysical origin discrimination networks}
	
	As detailed in the previous section. The preliminary templates obtained by the denoising models must be selected by the astrophysical origin discrimination network. The discrimination networks are deep convolutional neural networks for classification. For each preliminary template generated by a denoising network, an astrophysical origin discrimination network must be used to discriminate whether it is a template or not. Since 10 denoising networks are used, we need 10 astrophysical discrimination networks. 
	
	For every discrimination network, two classes of data are used for training (positive class and negative class). The positive class is the data with an astrophysical origin shape, and the negative class is the data without an astrophysical origin shape.
	
	Distinguishing whether the waveform has an astrophysical origin or not is a basic problem we face.  We employ automatic labeling for the positive class and manual labeling for the negative class. For each denoising network, 10 thousand samples of the positive class and 10 thousand samples of the negative class are labeled.
	
	Labeling the data with a positive class is simple. Enter the strain with GW signal $s(t)=h(t)+n(t)$ to the denoising model and obtain the output $\hat{h}(t)$. When the overlap of the $h(t)$ and $\hat{h}(t)$ is bigger than a threshold, we label $\hat{h}(t)$ as an astrophysical origin class (positive class). In this paper, we define the threshold as 0.8.
	
	The difficulty lies in labeling the data without astrophysical origin shape (negative class). We use three methods to generate the negative class for training. The three methods are described as follows.
	
	First, generate the negative samples from background noises (negative dataset I). Enter the strain with background noise $n(t)$  only to the denoising model and get the output $\hat{h}(t)$. The problem is that the output $\hat{h}(t)$ may have the astrophysical origin waveform shape. Thus the $\hat{h}(t)$ cannot be simply labeled as a negative class.  We solve this problem by manual labeling. Software for labeling one dimensional signals manually is developed. We use the software to manually screen the $\hat{h}(t)$ without astrophysical origin shape. 10,000 samples are generated for each denoising model by this method.
	
	Second, generate the negative samples from the training data containing GW signal (negative dataset II). Enter the strain with GW signal $s(t)=h(t)+n(t)$ to the denoising model and obtain the output $\hat{h}(t)$. The $\hat{h}(t)$ may not have the astrophysical origin shape. We use the overlap of the $h(t)$ and $\hat{h}(t)$ as the discriminant criteria for negative class labeling. When the overlap is below a threshold, we select $\hat{h}(t)$ as negative class. In this work, the threshold is set to 0.8. 10000 samples are generated by this method.
	
	Third, generate the negative samples from selected background noises that trigger the envelope extraction model (negative dataset III). In the proposed framework, since the envelope extraction process is a preprocessing stage of the framework. Screen the background noises from the first half month of August 2017 that trigger the envelope extraction model. Then put the screening samples into the denoising models. We select the output as the negative samples. 9080 samples are generated for each denoising model by this method.
	
	We train three kinds of astrophysical origin discrimination models (APOD\_MODEL\_I, APOD\_MODEL\_II, APOD\_MODEL\_III). The only difference among these three types of models is the selection of negative samples in the training dataset. The number of samples for each discrimination model’s positive and negative classes is balanced. The training datasets for the three models are shown in Table~\ref{tab:table3}.
	\begin{table*}[htbp]
		\centering
		\caption{\label{tab:table3}Training datasets for APOD\_MODEL\_I, APOD\_MODEL\_II, and APOD\_MODEL\_III.}
		\begin{ruledtabular}
			\begin{tabular}{ccccc}
				Name            &Negative dataset I&Negative dataset II&Negative dataset III&Samples in dataset \\
				\hline
				APOD\_MODEL\_I  &10000             &\dots              &\dots               &20000              \\
				APOD\_MODEL\_II &10000             &10000              &\dots               &40000              \\
				APOD\_MODEL\_III&10000             &10000              &9080                &58160              \\
			\end{tabular}
		\end{ruledtabular}
	\end{table*}
	
	\subsection{\label{sec:level2} Training scheme for the astrophysical origin discrimination model}
	
	In order to achieve the purpose of identifying the astrophysical origin signal, we use the astrophysical origin discrimination model described in the previous section to determine whether it is an astrophysical origin waveform. The loss function is binary cross-entropy, which is used to evaluate the deviation between the predicted values and the actual values in the training set. The gradient descent strategy is Adam, the learning rate is set to 1.5e-5, and the batch number value is 16.
	
	\section{\label{sec4} THE PERFORMANCE OF THE PROPOSED FRAMWORK}
	
	We have established the proposed framework for GW search. The framework takes the whitened strain as its input and provides a set with SNRs as output. This framework comprises an envelope extraction model, along with ten denoising models and ten models for discriminating astrophysical origins.  We have tested the performance of both the denoising models and the astrophysical discrimination models, as well as the overall framework, using both the test dataset and real strain data of Hanford and Livingston interferometers. The results convincingly showcase the effectiveness of the proposed framework.
	
	\subsection{\label{sec:level2} Results based on the test dataset}
	
	To test the framework’s performance, we generate 10 test datasets. The GPS time of the background noise is sampled from 1238904832 to 1239023616, which is different from the GPS time sampling employed in the training dataset.
	
	\subsubsection{Performance of the denoising models}
\begin{figure*}
		\includegraphics[scale=0.275]{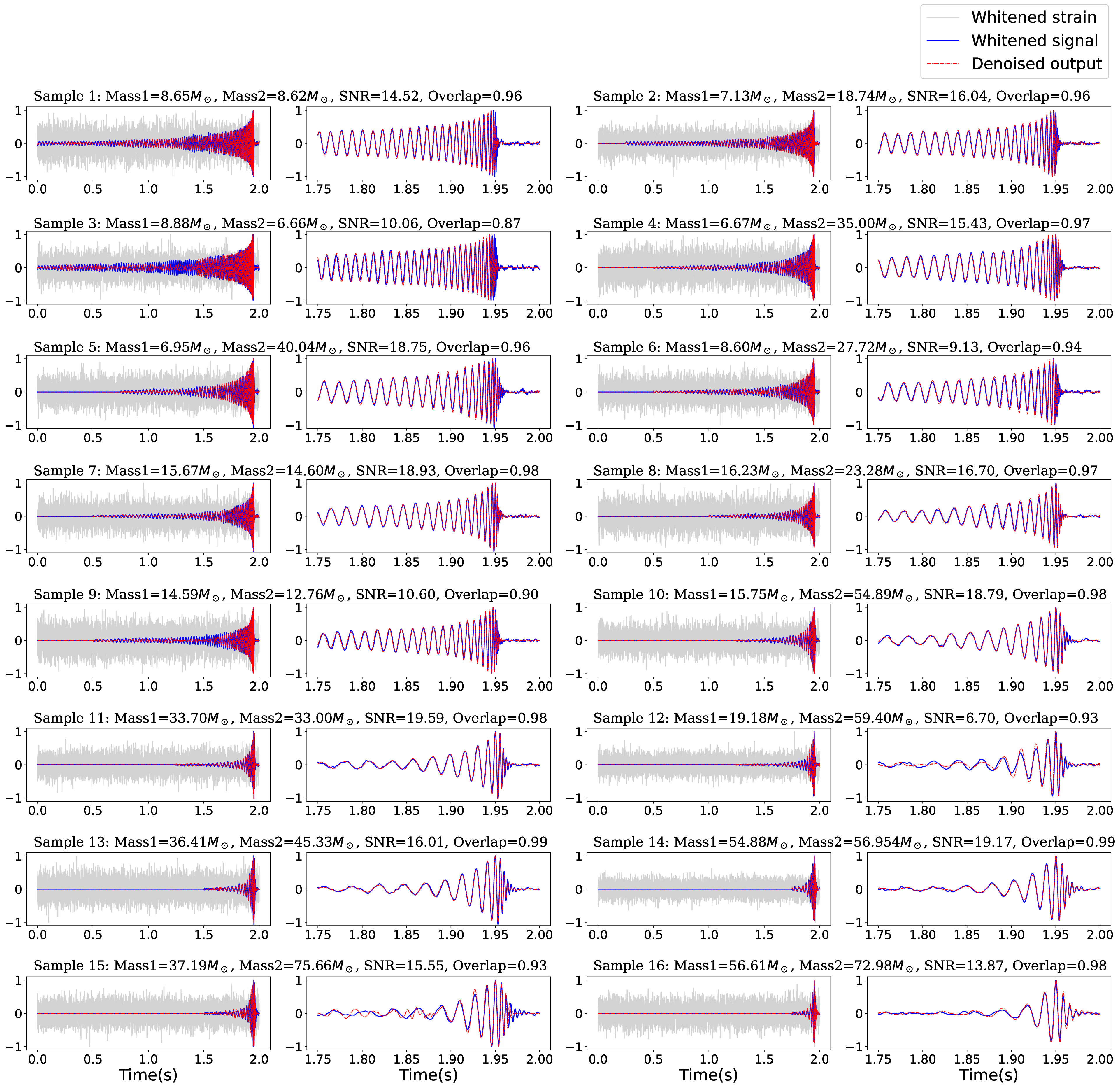}
		\caption{\label{fig:example_16} The whitened strain, whitened signal, and denoised output of 16 randomly selected samples in test dataset. The left column of every sample shows the whole 2 s time duration. The right column of every sample is the enlargement of the merger part including 0.25 s time duration.
		}
	\end{figure*}

	In this sub-section, we studied the performance of the denoising model based on the test dataset. We selected 16 samples from the test dataset randomly and put them into the corresponding denoising models. The whitened strain, the buried signal, and the output of the denoising model of every sample are shown in Fig.~\ref{fig:example_16}. The input and output lengths of the denoising model varies in accordance with the model’s associated parameter space. For instances with significant mass, the duration of the denoised output is less than 2 seconds; in such cases, the denoised output's residual portion is zero-padded to match a 2-second length. From Fig.~\ref{fig:example_16} we can see that in the majority of scenarios, the denoising model effectively recovered the buried signal. Notably, in the case of sample 3, the denoising model's output exhibits a precise recovery of the signal's merger phase, but with comparatively less efficacy for the early inspiral phase.
	
	\begin{figure*}
		\includegraphics[scale=0.35]{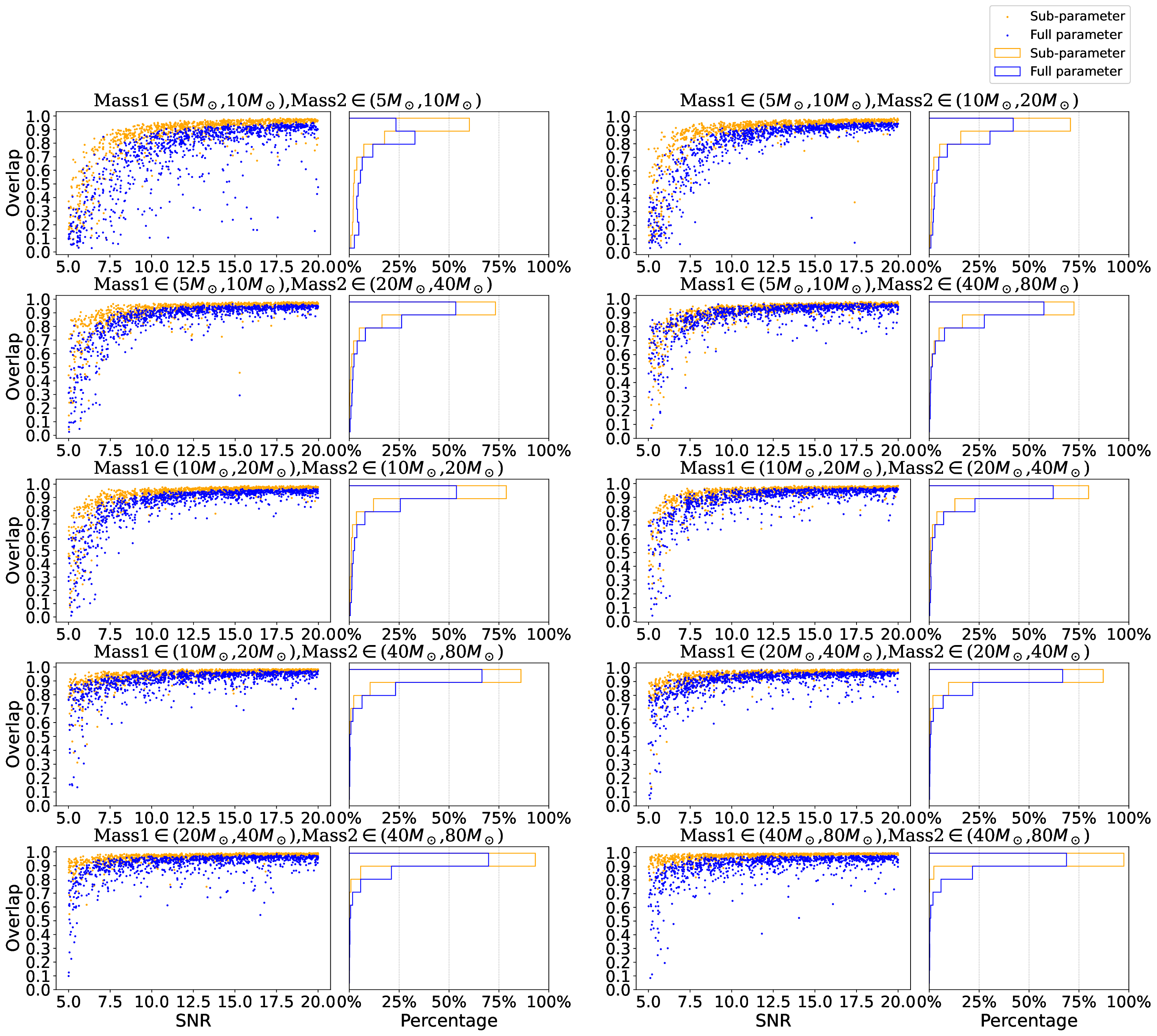}
		\caption{\label{fig:overlap} The overlap between the output of the denoising models and the buried GW signals. The denoising models trained by the sub-parameter space training data (orange color) and full parameter training data (blue color) are all investigated. Note that to avoid mutual coverage of a large number of points, only 1000 points of each model are shown in the scatterplot, and all the 10000 samples are used to plot the histogram in the right panel of each subplot.
		}
	\end{figure*}
	
	To measure the consistency of the denoised result and buried signal, the overlap between them is calculated. The overlap between the denoising model’s output $\hat{\boldsymbol{h}}$ and the buried signal $\boldsymbol{h}$ can be calculated by
	\begin{equation}
		o(\boldsymbol{h},\hat{\boldsymbol{h}})=\frac{\langle \boldsymbol{h},\hat{\boldsymbol{h}} \rangle}{\sqrt{\langle \boldsymbol{h},\boldsymbol{h}\rangle \langle \hat{\boldsymbol{h}},\hat{\boldsymbol{h}} \rangle}}.\label{eq13}
	\end{equation}
	
	The signal's duration within the sensitivity band of the detection network varies with the masses of the system. In the case of BBH systems, this duration spans from fractions of a second (for higher masses) to several seconds (for lower masses). 
	
	Other than the pioneer works \cite{44,45} for DL-based GW denoising, we take the variation of the duration with the source parameters into consideration, and the details were discussed in the previous section. We use the parameter division strategy, that the denoising models in the model bank are trained with a dataset corresponding to sub-parameter space. To illustrate the superiority of this parameter division, we established a denoising model trained with a full parameter dataset. Given that a 2 s length whitened signal adequately captures the characteristics of the signal originating from binary masses between $5\ M_\odot$ and $80\ M_\odot$, the input duration of the full parameter denoising model is set to 2 s. The model structure of the full parameter denoising model is the same as the model illustrated previously attended between $5\ M_\odot$ and $10\ M_\odot$.
	
	For each sample within every sub-parameter test dataset, we feed it into two distinct denoising models: the denoising model specific to the corresponding sub-parameter and the full-parameter denoising model. Subsequently, we obtain the respective outputs.
	
	We feed every sample in each sub-parameter test dataset into two denoising models (the corresponding sub-parameter denoising model and the full parameter denoising model). Subsequently, we obtain the respective outputs. Next, we calculate the overlap between the output and the buried GW signal, the outcome is presented in Fig.~\ref{fig:overlap}. The orange color denotes the sub-parameter denoising model’s result, and the blue color denotes the full parameter denoising model’s result.
	
	\begin{figure*}
		\includegraphics[scale=0.215]{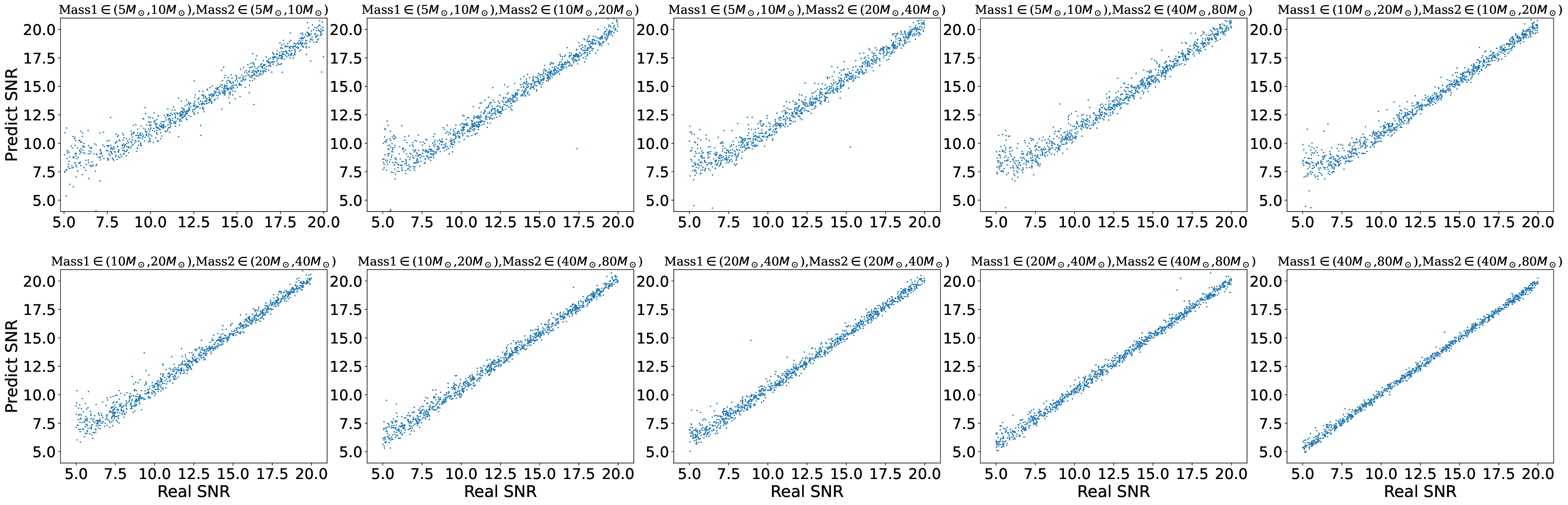}
		\caption{\label{fig:real_predict_snr} The SNRs of the ``noise+signal'' samples. Vertical coordinates indicate the SNRs calculated by matched filtering of the denoising output and the initial strain (our framework). Horizontal coordinates indicate the SNRs calculated by matched filtering of the buried signal and the initial strain. To avoid mutual coverage of a large number of points, each sub-figure only shows randomly selected 1000 points.
		}
	\end{figure*}
	
	Fig.~\ref{fig:overlap} highlights the superiority of the sub-parameter denoising model. We think the sub-parameter division introduced in this paper could also be integrated into the previously proposed denoising methods such as WaveNet and LSTM. The average overlap demonstrates an ascending trend corresponding to the binary masses. In the scenario of lowest mass systems, where $Mass1\in(5\ M_\odot,10\ M_\odot)$ and $Mass2\in(5\ M_\odot,10\ M_\odot)$, slightly more than 60\% of samples reproduce the true features of buried GW signals with overlaps greater than 0.9. For the case, both Mass1 and Mass2 fall within the range of $(40\ M_\odot,80\ M_\odot)$,  over 97.5\% of the sample outputs demonstrate overlaps surpassing 0.9. We will elucidate the rationale behind the model's diminished performance within the interval of low binary masses as follows.
	
	First, for the binary system with low-mass, the denoising method should be able to capture features reliant on long-range dependencies. For the low mass binary system case, the duration of the signal will be long. The U-Net employed in this study is grounded in convolutional neural networks (CNNs). However, CNN-based models exhibit a limitation in modeling long-range dependencies. Convolution operators are restricted to local receptive fields, and it's only after traversing multiple convolutional layers that the capability to perceive long-range dependencies is realized.
	
	The WaveNet tackled the challenge of local receptive fields by employing deeper models and using dilated convolutional layers \cite{44}. It's important to note that, for simplicity, the number of our denoising model’s parameters is much lower than that of WaveNet. The denoising models constitute just one component of our comprehensive framework, thus allowing for the integration of earlier denoising models like WaveNet \cite{44}, LSTM \cite{45}, and Waveformer \cite{46} into the framework. We will further investigate the denoising method for low-mass system in the future.
	
	\subsubsection{Performance of the SNR prediction}
	
	Given our utilization of SNR instead of $P_{score}$ to measure the confidence of samples containing GW signals, it's imperative to thoroughly examine the precision of SNR predictions by our framework. To this end, all samples in the test set are fed into the denoising model. We compute the matched filtering SNRs of the denoising output in the initial strain. A comparison between the SNRs calculated by our framework and the actual SNRs is illustrated in Fig.~\ref{fig:real_predict_snr}. Like the results presented in \cite{46}, the SNRs predicted by our framework exhibit consistency with the true SNRs.
	\begin{figure*}
		\includegraphics[scale=0.27]{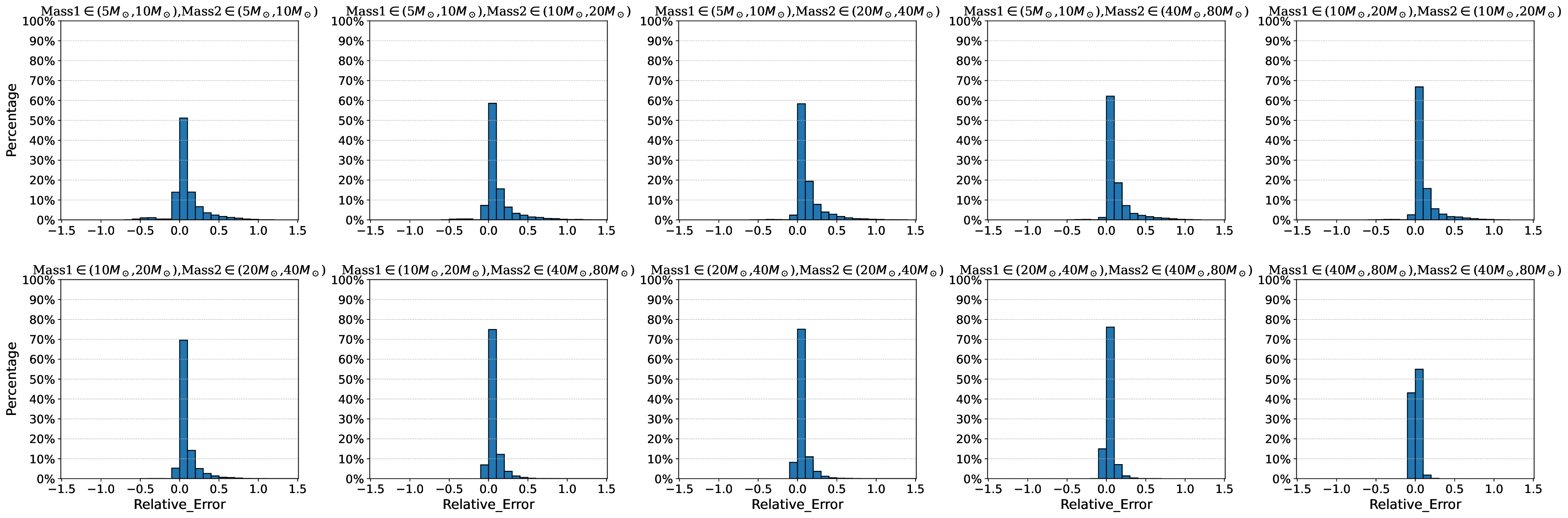}
		\caption{\label{fig:relative_erro} The distribution of the relative error range of the predicted SNR. We chose 0.1 as the relative error range’s step.
		}
	\end{figure*}
	
	Nevertheless, relying solely on Fig.~\ref{fig:real_predict_snr} isn't sufficient to prove the effectiveness of the proposed framework in GW search tasks. In GW search tasks, the method's capability to discriminate between ``noise" and ``noise+signal" is important. Fig.~\ref{fig:real_predict_snr} exclusively presents the results of denoising and matched filtering for scenarios involving ``noise+signal". At the end of this sub-section, we will investigate the output of the denoising and matched filtering result for ``noise" only instances.
	
	Fig.~\ref{fig:relative_erro} shows the distribution of the relative error of the predicted SNR over the test datasets. The relative error is calculated by
	\begin{equation}
		relative \_ error=\frac{SNR_{pred}-SNR_{true}}{|SNR_{true}|},\label{eq14}
	\end{equation}
	where $SNR_{pred}$ is the SNR predicted by the proposed framework and $SNR_{true}$ is the SNR of the buried template. We can see that the relative error is positive for most samples. The results indicate that most samples’ outputs are slightly bigger than the buried signal’s SNR. For the case $Mass1\in(5\ M_\odot,10\ M_\odot)$ and $Mass2\in(5\ M_\odot,10\ M_\odot)$, above 50\% samples’ relative errors are in the range [0,0.1). For the case  $Mass1\in(20\ M_\odot,40\ M_\odot)$ and $Mass2\in(20\ M_\odot,40\ M_\odot)$ above 70\% samples’ relative errors are in the range [0,0.1). 

	\begin{figure*}
		\includegraphics[scale=0.196]{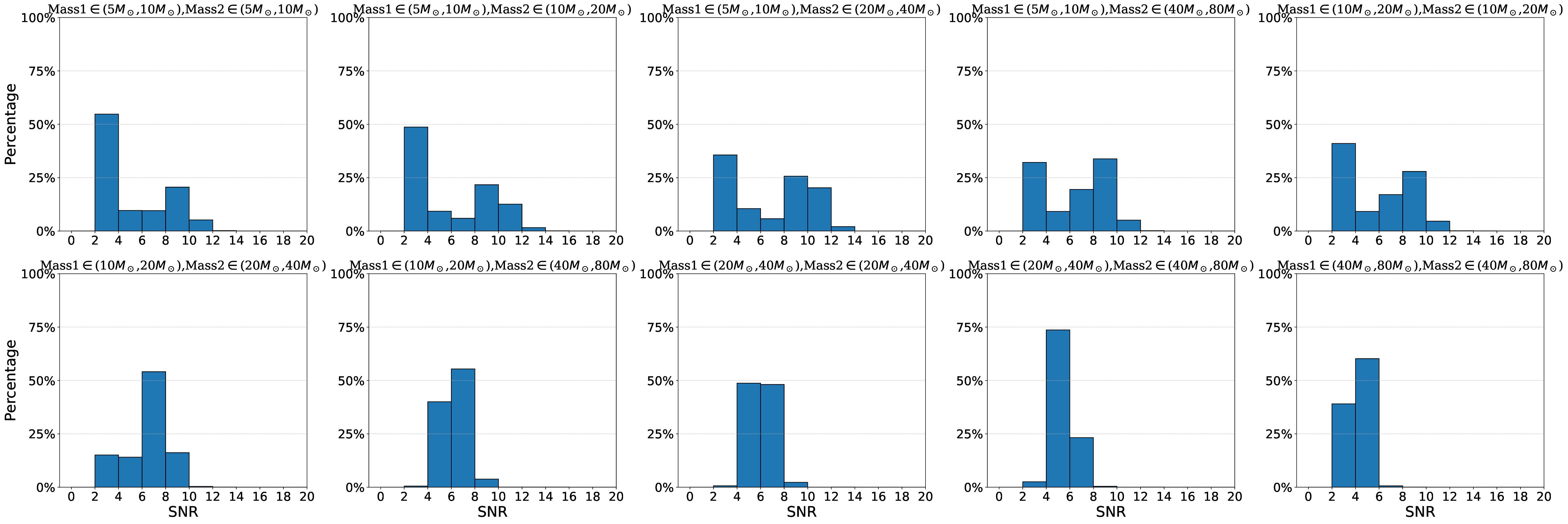}
		\caption{\label{noise_snr} Distribution of the matched filtering of denoised output of ``noise'' samples and the corresponding ``noise''.
		}
	\end{figure*}
	As previously explained, the proposed framework effectively estimates the SNR of the buried signal in cases involving ``noise+signal". Now, our focus shifts to the scenario of ``noise" alone. We applied the background noise from the 10 test sets to the denoising models and computed the matched filtering of the background noise and the denoising model's output. Each of the test sets encompasses 10,000 samples. The distributions of matched filtering SNRs for the denoising output are depicted in Fig.~\ref{noise_snr}.
	
	From Fig.~\ref{noise_snr} we can see that in the instance of lower mass cases (denoising model), a significant majority of samples exhibit denoising output's matched filtering SNRs lower than 12. Conversely, for models corresponding to higher mass systems such as when $Mass1\in(40\ M_\odot,80\ M_\odot)$ and $Mass2\in(40\ M_\odot,80\ M_\odot)$, the denoising output's SNR for the majority of samples tends to fall below 6.
	
	The results for the ``noise'' sample in Fig.~\ref{noise_snr} indicate that the denoise+MF only cannot be used for GW search. This is due to the presence of numerous instances where noise only yields relatively higher matched filtering SNRs, leading to potential misclassification of "noise" samples as ``noise+signal" instances. As mentioned in Section II, after denoising we use astrophysical discrimination networks to discriminate if the denoising output exhibits a shape of astrophysical origin or not. In the subsequent sub-section, we will delve into the performance of the astrophysical origin discrimination networks in the test set.

	\subsubsection{Performance of the astrophysical origin discrimination networks}
	
	From the previous sub-section, we have learned that only the denoising and matched filtering process is inadequate for the GW search task. To solve this problem, we introduce an astrophysical origin discrimination network. As described in Sec. III C, for each sub-parameter space three kinds of astrophysical discrimination networks are constructed (APOD\_MODEL\_I, APOD\_MODEL\_II, APOD\_MODEL\_III) based on the training datasets. In this sub-section, we will investigate the performance of the astrophysical origin discrimination networks.
	
\begin{figure*}
		\includegraphics[scale=0.196]{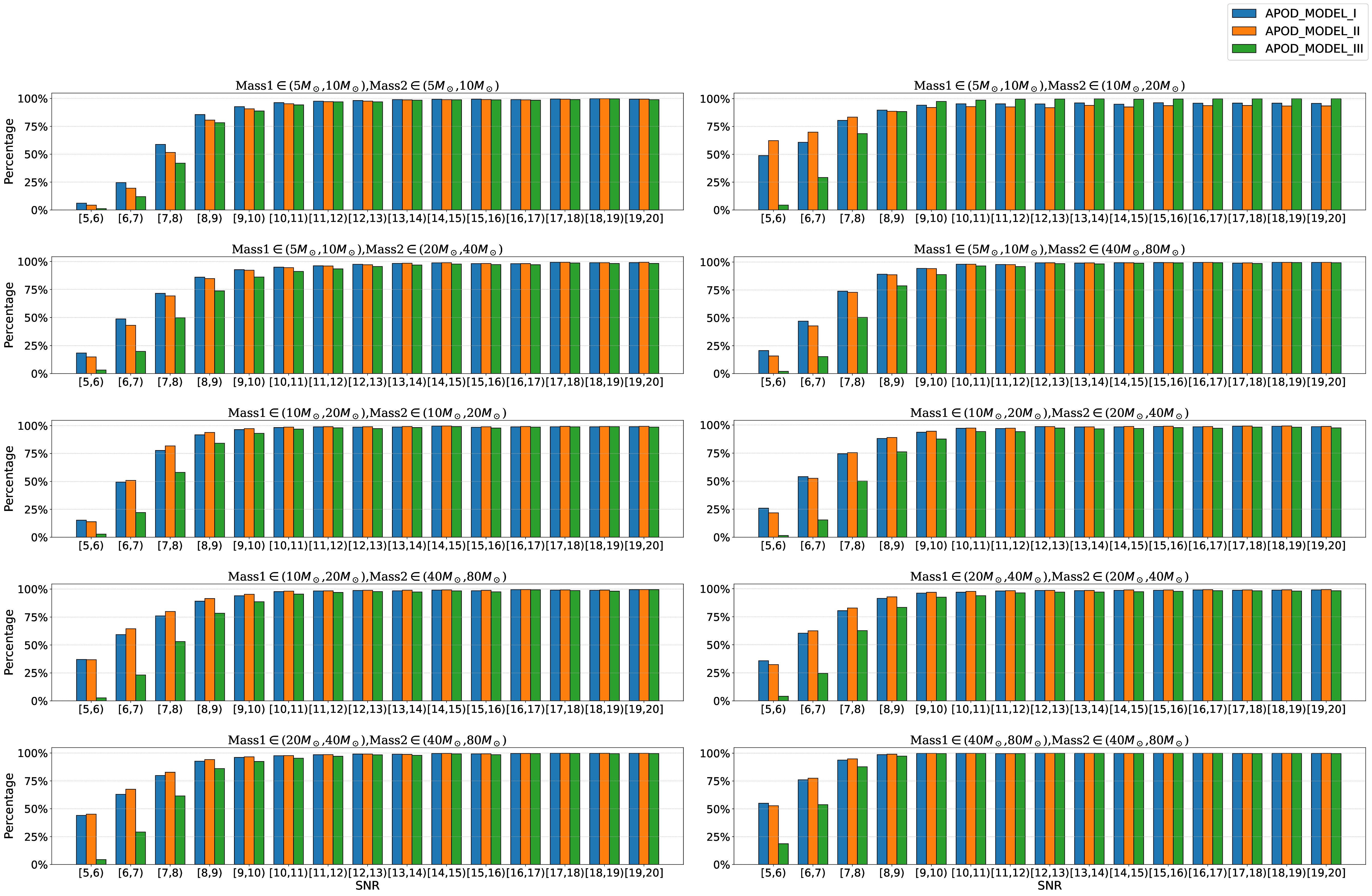}
		\caption{\label{fig:screening_percentage_3model} The percentage of correctly classified the denoised results from ``signal + noise''.
		}
	\end{figure*}

	We put the denoising outputs of “signal+noise” into astrophysical origin discrimination networks and get the corresponding classification outcomes. A statistical analysis of the classification results is shown in Fig.~\ref{fig:screening_percentage_3model}. We can see that in the low SNR range case, the APOD\_MODEL\_I and APOD\_MODEL\_II outperform the APOD\_MODEL\_III. In the high SNR case, the three kinds of astrophysical origin discrimination models perform similarly. For APOD\_MODEL\_III, a significant majority of the denoising outputs can be correctly classified when the buried signal’s $SNR \geq 8$. Within the SNR range of 7 to 8, approximately half of the results are correctly classified, while for SNRs falling below 7, most of the outcomes are misclassified. This is in consistent with the results in Fig.~\ref{fig:overlap} and Fig.~\ref{fig:real_predict_snr}.
	
	\subsubsection{Performance of the GW search framework with PTP, TS, and MF stages}
	
	In this sub-section, we will investigate the performance of the proposed GW search framework by test dataset. The test set is composed of samples with relatively short durations. Given that the ``noise" samples often fail to trigger the STP stage, this specific stage won't be employed in the experiments of this section. The comprehensive framework including STP will undergo testing using the confident events and half-month detected strain data detailed in Sections IV B and IV C.

	Similar to various other classification tasks, we use receiver operator characteristic (ROC) curves to evaluate the performance of the search methods. The ROC curve reflects the relationship between true alarm probability (TAP) and false alarm probability (FAP). The TAP and FPR are defined as follows.
	\begin{equation}
		TAP:=\frac{TP}{TP+FN},\label{eq15}
	\end{equation}
	\begin{equation}
		FAP:=\frac{FP}{FP+TN}.\label{eq16}
	\end{equation}
	
	In this context, TP represents true positives, FP signifies false positives, TN corresponds to true negatives, and FN denotes false negatives. In the proposed framework, two comparable metrics are used. One is the output of the astrophysical discrimination network $(\boldsymbol{o}_{astro})$ and the other is the output of the matched filtering process (SNR). For simplification, we set the threshold of $\boldsymbol{o}_{astro}$ to 0.5. In the case of $\boldsymbol{o}_{astro}<0.5$, the proposed framework categorizes the strain as noise. Conversely, when $\boldsymbol{o}_{astro}>0.5$, the MF stage will be further conducted and the confidence of the strain containing signal is evaluated using SNR. When the SNR surpasses a predefined threshold, a GW trigger will generate. Sweeps the SNR threshold from 5 to 20, recording the TAP and FAP for each threshold value to produce the ROC curve of the proposed framework. It's important to acknowledge that due to the fixed threshold value of  $\boldsymbol{o}_{astro}$, certain "noise+signal" samples might not surpass this threshold, thereby preventing TAP from reaching 100\% in such cases.
	\begin{figure*}
		\includegraphics[scale=0.193]{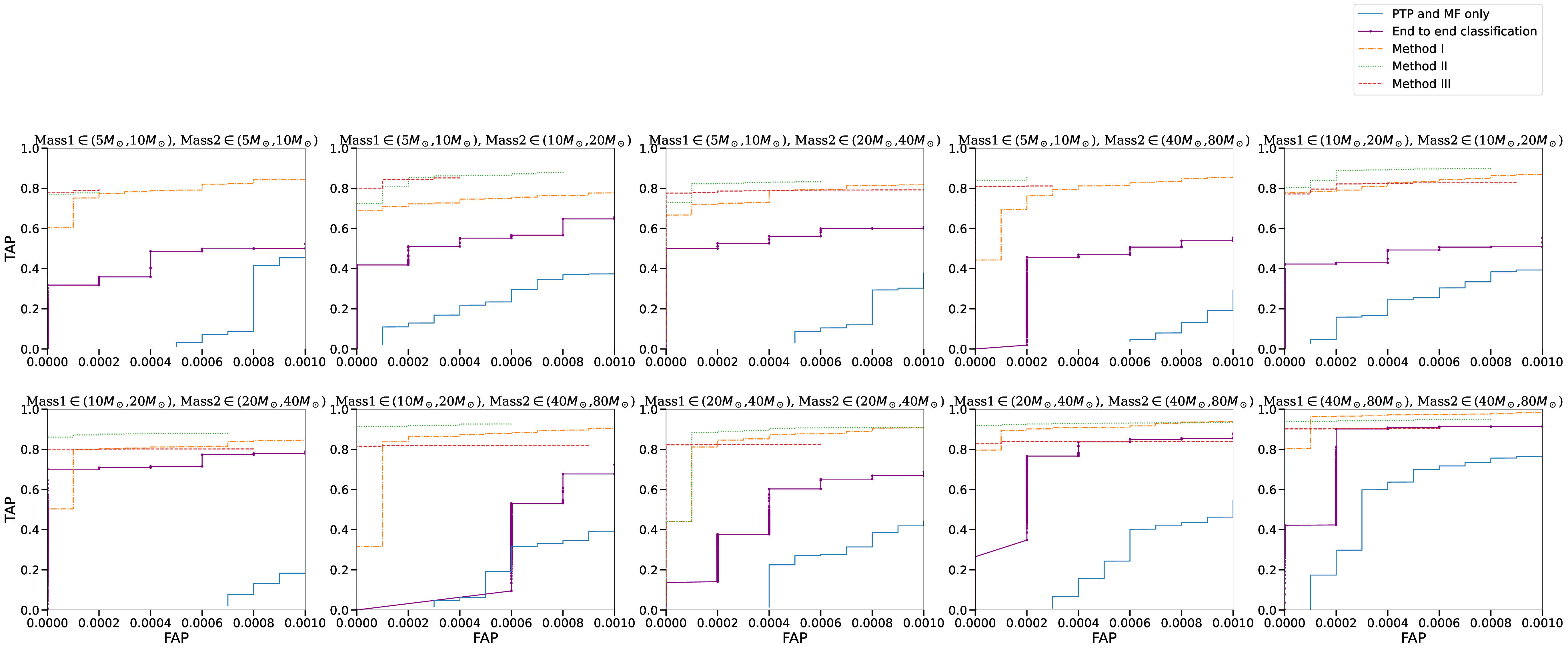}
		\caption{\label{fig:roc} The ROC curves of the proposed framework (Method I, II, and III), the framework without the TS stage (PTP and TS stages only), and the end-to-end DL method (DL classification).
		}
	\end{figure*}
	
	Fig.~\ref{fig:roc} shows the ROC curves of the proposed the GW search method. For comparison, the results of GW search method based on PTP and MF only are shown too. We also constructed 10 end-to-end GW classification models. Different from the classification model in other studies, every classification model only attends to the GW morphology of a specific sub-parameter space. The classification model takes the whitened strain as input and generates an output in the form of $P_{score}$, which falls within the range of 0 to 1.
	
	We conduct three methods (Method I, II, and III) based on PTP, TS, and MF stages. The disparity among these three methods lies in the choice of astrophysical origin discrimination networks during the TS stage. Specifically, we employ APOD\_MODEL\_I for Method I, APOD\_MODEL\_II for Method II, and APOD\_MODEL\_III for Method III.
	
	For the real data, most of the strain's GW signals are often indiscernible and are assumed to consist mainly of background noise. Despite a search method having a relatively low FAP, numerous false triggers can still emerge, especially in scenarios involving long-duration detection. Therefore, it is only making sense for the low FAP range. Fig.~\ref{fig:roc} only displays the FAP range below $10^{-3}$.
	
	From Fig.~\ref{fig:roc} we can see the superiority of the GW search frameworks based on the PTP, TS, and MF stages. In the context of test datasets, the performances of Method I, II, and III exhibit similarities. Specifically, when $FAP=10^{-4}$, for most cases of the proposed framework the TAPs are consistently around 0.8. On the contrary, for the framework without the TS stage, when FAP=10-4, the TAPs are either 0 or exceedingly close to 0. This is consistent with Fig.~\ref{noise_snr} and proves once again the importance of the astrophysical origin discrimination model. The matched filtering of the “noise” sample’s denoising output and the sample may be a large value, resulting in an output SNR that does not resemble the SNR of a GW-like waveform.
	
	Interestingly, for most cases in Fig.~\ref{fig:roc}, the FAP of the end-to-end GW classification models can be effectively brought down to the magnitude of $10^{-4}$. The results in \cite{30} and \cite{28} all show that the FAP of the model proposed by Gabbard \cite{21} is hardly reduced to $10^{-4}$. This suggests that an end-to-end GW search model concentrated on a sub-parameter space has the potential to significantly enhance search performance. Within the context of sub-parameter GW search, the ability to optimize the duration of the time window for data analysis becomes feasible. This shows the potential of the sub-parameter end-to-end classification model for GW search, which we believe is a promising avenue for further investigation in the future.
	\begin{table*}[htbp]
		\centering 
		\caption{SNRs of the GW events reported by LVK in O1, O2, and O3 \cite{1,2,3}. Events that only both Hanford and Livingston interferometers work well are listed. The SNRs predicted by the pipelines cWB, GstLAL, MBTA, PyCBC, and PyCBC\_BBH are shown. Here we also show SNRs predicted by our proposed framework. Because three kinds of astrophysical origin discrimination networks are constructed, we use three GW search methods (Method I, Method II, and Method III). Method I, II, and III use APOD\_MODEL\_I, APOD\_MODEL\_II, and APOD\_MODEL\_III for the astrophysical origin discrimination network, respectively. Note that some events’ FAR or $p_{astro}$ are not in the confident event range, and, in this case, the SNR value is not shown. The time differences between the peak value of the denoised strain (Method II) of Hanford and Livingston interferometers are also shown.} 
		\begin{ruledtabular}
			\begin{tabular}{cccccccccc}
				Name           & cWB  & GstLAL & MBTA & PyCBC & PyCBC BBH &Method I & Method II & Method III & \makecell{Time differences \\ of Method II(s)}   \\
				\hline
				GW150914       & 25.2 & 24.4   &   -  & 23.6  &    -      & 24.5    & 24.5      & 24.4       & 0.00439  \\
				GW151012       & \dots& 10.0   &   -  & 9.5   &    -      & 13.4    & 10.6      & \dots      & 0.00293  \\
				GW151226       & 11.39& 13.1   &   -  & 13.1  &    -      & 14.8    & 13.7      & \dots      & 0.00269  \\
				GW170104       & 13.0 & 13.0   &   -  & 13.0  &    -      & 14.4    & 14.4      & 14.2       & 0.00024  \\
				GW170608       & 14.1 & 14.9   &   -  & 15.4  &    -      & 17.0    & 16.9      & 16.9       & 0.00024  \\
				GW170729       & 10.2 & 10.8   &   -  & 9.8   &    -      & 10.9    & 11.6      & 10.8       & 0.00586  \\
				GW170809       & \dots& 12.4   &   -  & 12.2  &    -      & 14.7    & 13.4      & 13.4       & 0.01220  \\
				GW170814       & 17.2 & 15.9   &   -  & 16.3  &    -      & 17.1    & 17.1      & 17.1       & 0.01220  \\
				GW170818       & \dots& 11.3   &   -  & \dots &    -      & \dots   & \dots     & \dots      & \dots    \\
				GW170823       & 10.8 & 11.5   &   -  & 11.1  &    -      & 12.7    & 12.7      & \dots      & 0.00537  \\
				GW190403-051519&\dots &\dots   &\dots & 16.3  &    8.0    & \dots   & \dots     & \dots      & \dots    \\
				GW190408-181802&14.8  &14.7    &14.4  & 13.1  &    13.7   & 15.6    & 15.6      & 15.6       & 0.00659  \\
				GW190412       &19.7  &19.0    &18.2  & 17.4  &    17.9   & 20.0    & 20.0      & 19.5       & 0.00122  \\
				GW190413-052954&\dots &\dots   &\dots &\dots  &    8.5	  &9.1 	    &9.1 	    & \dots      &0.00049   \\
				GW190413-134308&\dots &\dots   &10.3  &\dots  &    8.9	  &10.3     &9.9 	    & \dots      &0.00098   \\
				GW190421-213856&9.3   &10.5    &9.7   &10.1   &    10.1	  &11.2     &11.2 	    & 10.2       &0.00171   \\
				GW190426-190642&\dots &\dots   &\dots &\dots  &    9.6	  &\dots    &\dots 	    & \dots      &\dots     \\
				GW190503-185404&11.5  &12.0    &12.8  &12.2   &    12.2	  &14.9     &14.9  	    & 12.4       &0.00610   \\
				GW190512-180714&10.7  &12.2    &11.7  &12.4   &    12.4	  &13.4     &\dots 	    & \dots      &\dots     \\
				GW190513-205428&\dots &12.3    &13.0  &\dots  &    11.8	  &14.1     &14.1 	    & 13.0       &0.0000    \\
				GW190514-065416&\dots &\dots   &\dots &\dots  &    8.4	  &9.1      &\dots 	    & \dots      &\dots     \\
				GW190517-055101&10.7  &10.8    &11.3  &10.4   &    10.3	  &10.9     &11.1 	    & 10.9       &0.00342   \\
				GW190519-153544&14.0  &12.4    &13.7  &13.2   &    13.2	  &14.8     &15.1 	    & 14.8       &0.00269   \\
				GW190521       &14.4  &13.3    &13.0  &13.7   &    13.6	  &\dots    &\dots 	    & \dots      &\dots     \\
				GW190521-074359&24.7  &24.4    &22.2  &24.0   &    24.0	  &25.0     &24.4 	    & 24.4       &0.00098   \\
				GW190527-092055&\dots &8.7     &\dots &\dots  &    \dots  &8.8      &\dots 	    & \dots      &\dots     \\
				GW190602-175927&11.1  &12.3    &12.6  &11.9   &    11.9	  &12.0     &12.8 	    & 11.9       &0.01293   \\
				GW190701-203306&10.2  &11.7    &11.3  &11.9   &    11.7	  &10.7     &10.7 	    & 10.4       &0.00513   \\
				GW190706-222641&12.7  &12.5    &11.9  &11.7   &    12.6	  &13.6     &13.6 	    & 12.7       &0.00488   \\
				GW190707-093326&\dots &13.2    &12.6  &13.0   &    13.0	  &14.9     &14.2 	    & 14.2       &0.00708   \\
				GW190719-215514&\dots &\dots   &\dots &\dots  &    8.0	  &10.1     &10.1 	    & \dots      &0.00269   \\
				GW190720-000836&\dots &11.5    &11.6  &10.6   &    11.4	  &\dots    &\dots 	    & \dots      &\dots     \\
				GW190725-174728&\dots &\dots   &9.8   &9.1    &    8.8	  &12.2     &\dots 	    & \dots      &\dots     \\
				GW190727-060333&11.4  &12.1    &12.0  &11.4   &    11.1	  &13.6     &12.6 	    & 12.0       &0.00537   \\
				GW190728-064510&\dots &13.4    &13.1  &13.0   &    13.0	  &14.9     &14.9 	    & 14.9       &0.00220   \\
				GW190731-140936&\dots &8.5     &9.1   &\dots  &    7.8	  &9.5      &8.6 	    & \dots      &0.01050   \\
				GW190803-022701&\dots &9.1     &9.0   &\dots  &    8.7	  &10.1     &9.6 	    & \dots      &0.00220   \\
				GW190805-211137&\dots &\dots   &\dots &\dots  &    8.3	  &8.8      &\dots 	    & \dots      &\dots     \\
				GW190814       &\dots &22.2    &20.4  &19.5   &    \dots  &23.7     &23.4 	    & 23.4       &0.00806   \\
				GW190828-063405&16.6  &16.3    &15.2  &13.9   &    15.9	  &17.3     &17.3 	    & 17.3       &0.00317   \\
				GW190828-065509&\dots &11.1    &10.8  &10.5   &    10.5	  &12.3     &11.3 	    & \dots      &0.00024   \\
				GW190915-235702&12.3  &13.0    &12.7  &13.0   &    13.1	  &14.1     &13.6 	    & 13.6       &0.00342   \\
				GW190916-200658&\dots &\dots   &8.2   &\dots  &    7.9	  &9.2      &9.2  	    & \dots      &0.0.051   \\
				GW190917-114630&\dots &9.5     &\dots &\dots  &    \dots  &\dots    &\dots 	    & \dots      &\dots     \\
				GW190924-021846&\dots &13.0    &11.9  &12.4   &    12.5	  &\dots    &\dots 	    & \dots      &\dots     \\
				GW190926-050336&\dots &9.0     &\dots &\dots  &    \dots  &9.0      &9.0  	    & \dots      &0.00342   \\
				GW190929-012149&\dots &10.1    &10.3  &\dots  &    \dots  &\dots    &\dots 	    & \dots      &\dots     \\
				GW190930-133541&\dots &10.1    &10.0  &9.8    &    10.0	  &\dots    &\dots 	    & \dots      &\dots     \\
				GW191103-012549&\dots &\dots   &\dots &9.3    &    9.3	  &\dots    &\dots 	    & \dots      &\dots     \\
				GW191105-143521&\dots &\dots   &10.7  &9.8    &    9.8	  &11.9     &\dots 	    & \dots      &\dots     \\
				GW191109-010717&15.6  &15.8    &15.2  &13.2   &    14.4	  &16.6     &15.8 	    & 15.8       &0.00293   \\	
				GW191113-071753&\dots &\dots   &9.2   &\dots  &    \dots  &\dots    &\dots 	    & \dots      &\dots     \\			
			\end{tabular}
		\end{ruledtabular}
		\label{tab:label4}%
	\end{table*}
	\begin{table*}[htbp]
		\begin{ruledtabular}
			\begin{tabular}{cccccccccc}
				Name           & cWB  & GstLAL & MBTA & PyCBC & PyCBC BBH &Method I & Method II & Method III & \makecell{Time differences \\ of Method II(s)}   \\
				\hline
				GW191126-115259&\dots &\dots   &\dots &\dots  &    8.5	  &\dots    &\dots 	    & \dots      &\dots     \\
				GW191127-050227&\dots &10.3    &9.8   &\dots  &    8.7	  &9.5      &9.5  	    & \dots      &0.00708   \\
				GW191129-134029&\dots &13.3    &12.7  &12.9   &    12.9	  &\dots    &\dots 	    & \dots      &\dots     \\
				GW191204-110529&\dots &\dots   &\dots &\dots  &    8.9	  &10.5     &10.5 	    & \dots      &0.00244   \\
				GW191204-171526&17.1  &15.6    &17.1  &16.9   &    16.9	  &17.3     &17.3 	    & 17.3       &0.00171   \\					
				GW191215-223052&9.8   &10.9    &10.8  &10.3   &    10.2	  &10.8     &11.3 	    & \dots      &0.00635   \\
				GW191219-163120&\dots &\dots   &\dots &8.9    &    \dots  &\dots    &\dots 	    & \dots      &\dots     \\
				GW191222-033537&11.1  &12.0    &10.8  &11.5   &    11.5	  &14.0     &11.8  	    & 11.8       &0.00366   \\
				GW191230-180458&10.3  &10.3    &\dots &\dots  &    9.9	  &12.1     &10.5 	    & 9.9        &0.00195   \\
				GW200115-042309&\dots &11.5    &11.2  &10.8   &    \dots  &\dots    &\dots 	    & \dots      &\dots     \\
				GW200128-022011&8.8   &10.1    &9.4   &9.8    &    9.9	  &10.1     &10.1 	    & 9.8        &0.00317   \\
				GW200129-065458&\dots &26.5    &\dots &16.3s  &    16.2	  &27.3     &26.9 	    & 26.5       &0.00269   \\
				GW200202-154313&\dots &11.3    &\dots &\dots  &    10.8	  &\dots    &\dots 	    & \dots      &\dots     \\
				GW200208-130117&\dots &10.7    &10.4  &9.6    &    10.8	  &10.6     &\dots 	    & \dots      &\dots     \\
				GW200208-222617&\dots &\dots   &\dots &\dots  &    7.9	  &\dots    &\dots 	    & \dots      &\dots     \\
				GW200209-085452&\dots &10.0    &9.7   &\dots  &    9.2	  &9.5      &\dots 	    & \dots      &\dots     \\
				GW200210-092254&\dots &9.5     &\dots &8.9    &    8.9	  &\dots    &\dots 	    & \dots      &\dots     \\
				GW200216-220804&\dots &9.4     &\dots &\dots  &    8.7	  &8.8      &8.8  	    & 8.8        &0.01221   \\
				GW200219-094415&9.7   &10.7    &10.6  &9.9    &    10.0	  &12.7     &11.6 	    & \dots      &0.00122   \\
				GW200220-061928&\dots &\dots   &\dots &\dots  &    7.5	  &\dots    &\dots 	    & \dots      &\dots     \\
				GW200220-124850&\dots &\dots   &8.2   &\dots  &    \dots  &10.2     &\dots 	    & \dots      &\dots     \\
				GW200224-222234&18.8  &18.9    &19.0  &19.2   &    18.6	  &19.6     &19.6 	    & 18.3       &0.00098   \\
				GW200225-060421&13.1  &12.9    &12.5  &12.3   &    12.3	  &14.5     &14.5 	    & 14.5       &0.00684   \\
				GW200306-093714&\dots &\dots   &8.5   &\dots  &    \dots  &\dots    &\dots 	    & \dots      &\dots     \\
				GW200308-173609&\dots &\dots   &\dots &\dots  &    8.0	  &\dots    &\dots 	    & \dots      &\dots     \\
				GW200311-115853&16.2  &17.7    &16.5  &17.0   &    17.4	  &18.4     &16.7 	    & 16.6       &0.00342   \\
				GW200316-215756&\dots &10.1    &\dots &9.3    &    9.3	  &\dots    &\dots 	    & \dots      &\dots     \\
				GW200322-091133&\dots &\dots   &9.0   &\dots  &    9.6	  &\dots    &\dots 	    & \dots      &\dots     \\
				\hline
				Detection(\%)  & 40.0 &73.8    &60.0  &62.5   &    73.8   &71.3     &60.0       &41.3        &          \\
			\end{tabular}
		\end{ruledtabular}	
	\end{table*}
	
	\subsection{\label{sec:level2} Results for confident events}
	
	In this sub-section, we investigate the detection performance of the proposed framework on the confident events listed in the GWTC-1, GWTC-2, GWTC-2.1, and GWTC-3 catalogs. We first investigate the denoising performance related to the confident events. Subsequently, all events that both Hanford and Livingston interferometers work well are studied by all the four stages (STP, PTP, TS, and MF) of the proposed framework. 
	
	Here we analyze the denoising outcomes in relation to strains containing confident events. To begin, we extracted the envelope (STP stage detailed in Sec.II) of the whitened strain. Subsequently, we cut $T_i\ s$ whitened strain to the \emph{i}-th denoising model and get the output. Note that the envelope’s peak time corresponds to the final 0.05 s of the $T_i\ s$ data. We selected 16 samples of the confident events from the Hanford interferometer. Our investigation encompasses a wide range of masses for the detected binary black hole (BBH) mergers.
	
\begin{figure*}
		\includegraphics[scale=0.27]{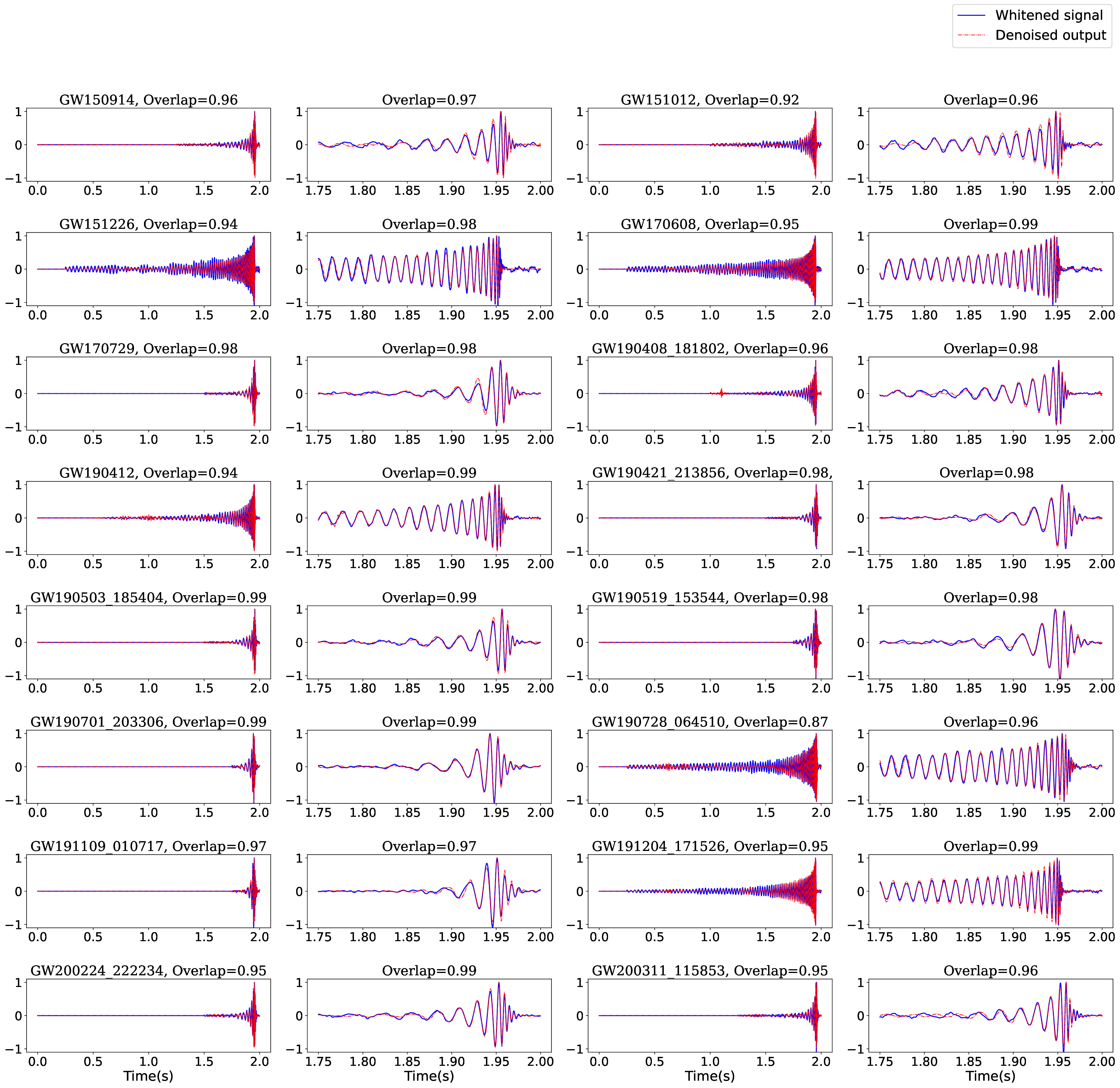}
		\caption{\label{true_events_denoise} Comparison between denoised signals from real events and their optimal templates. 	
		}
	\end{figure*}

	The comparison between the denoising outputs and the optimal templates (sampled from the posterior distribution provided by GWTC-2.1 and GWTC-3) is depicted in Fig.~\ref{true_events_denoise}. Fig.~\ref{true_events_denoise} illustrates that our denoising model effectively retrieves the signal from the confident events. Intriguingly, our denoising model successfully reconstructed the GW170608 signal detected by the Hanford interferometer. The overlap between the denoising output and the template is notably high, measuring 0.95 at a 2-second time scale and reaching an impressive 0.99 at a 0.25-second time scale. Notably, the denoising output of GW170608 outperforms the WaveNet denoising result, which achieved an overlap of 0.73 according to previous findings \cite{44}.
	
	In order to comprehensively evaluate the effectiveness of the entire GW search framework on real events, we further conducted experiments. All events' whitened strains were fed into the framework encompassing the STP, PTP, TS, and MF stages. For the TS stage, we employed three distinct models: APOD\_MODEL\_I, APOD\_MODEL\_II, and APOD\_MODEL\_III, corresponding respectively to Method I, II, and III.
	
	Table~\ref{tab:label4} shows the SNRs predicted by our framework (Method I, II, and III) alongside those generated by five famous GW search pipelines (cWB, GstLAL, MBTA, PyCBC, and PyCBC\_BBH). Notably, GWTC-1 didn’t present MBTA and PyCBC\_BBH’s results, so the 10 O1 and O2 BBH events’ network SNRs predicted by the two pipelines were not shown.
	
	Once the strains from both the Hanford and Livingston interferometers within the same sub-parameter space (denoising and astrophysical origin discrimination) successfully pass the STP, PTP, and TS stages, the MF stage is subsequently engaged to compute the network SNR of that sub-parameter space. For each method, the outcome takes the form of a network SNR set.
	
	Every SNR in the set corresponds to the SNR associated with a distinct sub-parameter space. The final network SNR is calculated by
	\begin{equation}
		NetworkSNR=\mathop{\max}\limits_{i}NetworkSNR_i,\label{eq17}
	\end{equation}
	where $NetworkSNR_i$ is the predicted network SNR, and
	\begin{equation}
		NetworkSNR_i=\sqrt{SNR_{i,H}^2+SNR_{i,L}^2},\label{eq18}
	\end{equation}
	where $SNR_{i,H}$ and $SNR_{i,L}$ are the predicted SNR values for the \emph{i}-th sub-parameter space in Hanford and Livingston, respectively.
	
	In O1 and O2, 9 confident events are successfully identified using both Method I and Method II. However, owing to the weakness of the signal at the Hanford interferometer, all three methods fail to detect GW170818. Method III also fails to successfully detect GW151012, GW151226, and GW170823. Method I achieves a success rate of approximately 71\% in event detection, followed by Method II at 60\%, and Method III at 41\%. Note that this does not mean that the superior of Method I to MBTA and PyCBC pipelines because the detection benchmarks of our proposed framework and traditional MF-based pipelines are different. Conventional MF-based pipelines employ $p_{astro}$ as their detection benchmark, while we utilize the output of the astrophysical origin discrimination network, $\boldsymbol{o}_{astro}$, for our detection benchmark. 
	
	In the case of Method II, we also examined the merge time differences in denoised strains between the Hanford and Livingston interferometers. The merge time difference is defined by
	\begin{equation}
		diff \_ t=|\mathop{\arg \max}\limits_{t}\hat{h}_L(t)-\mathop{\arg \max}\limits_{t}\hat{h}_H(t)|,\label{eq19}
	\end{equation}
	where $\hat{h}_H(t)$ is the denoised strain of Hanford, and $\hat{h}_L(t)$ is the denoised strain of Livingston. The results (presented in the final column of Tabel~\ref{tab:label4}) show that the merge time differences of all identified event mergers are within 13 ms. Similar to the MF-based approach, the time differences across multiple interferometers can be employed to further validate the results.
	\begin{figure*}
		\includegraphics[scale=0.5]{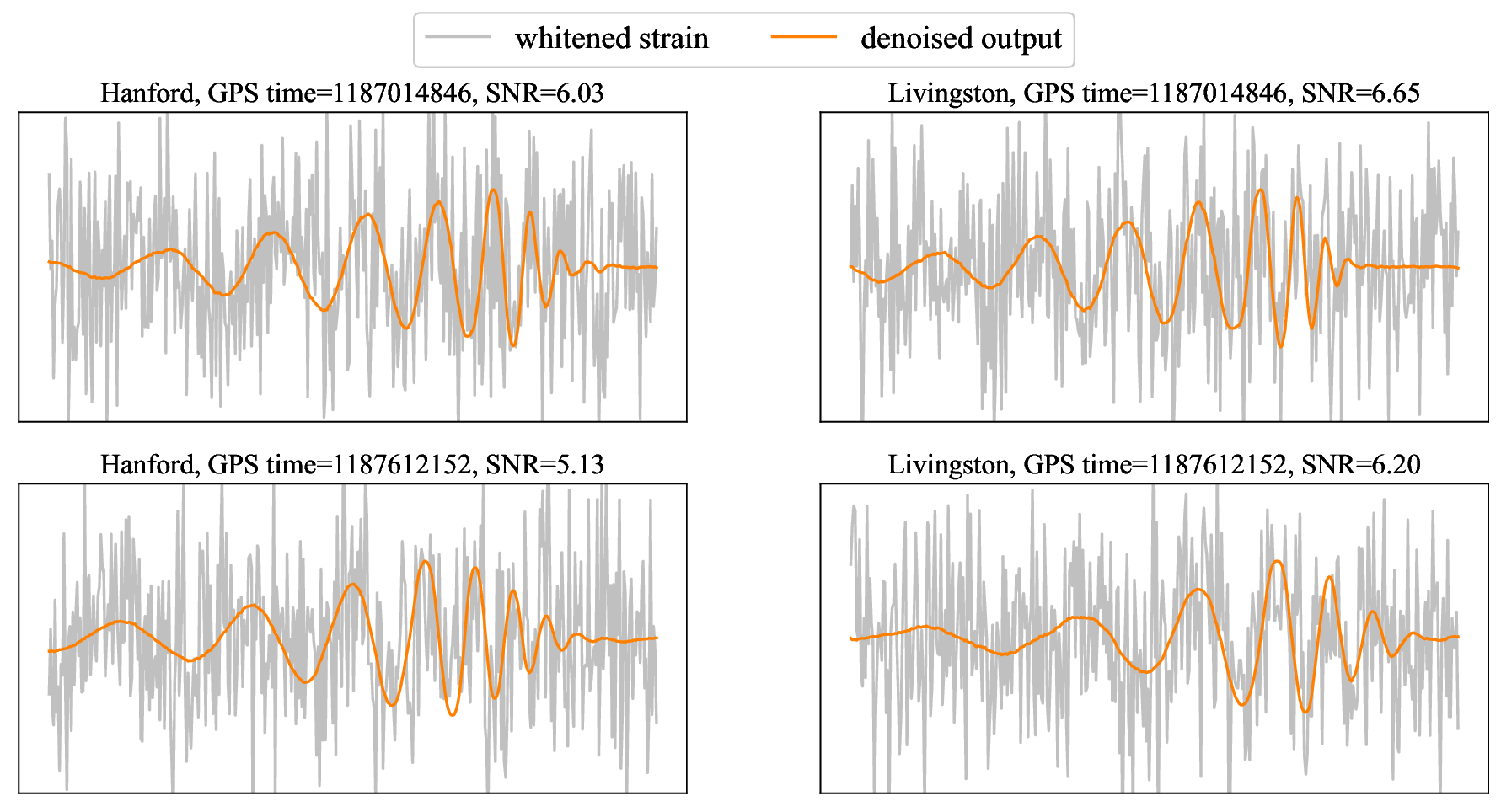}
		\caption{\label{strainCompare} Whitened strain and denoised outputs near the two false triggers of Method II.}
    \end{figure*}
    \begin{figure*}
		\includegraphics[scale=0.5]{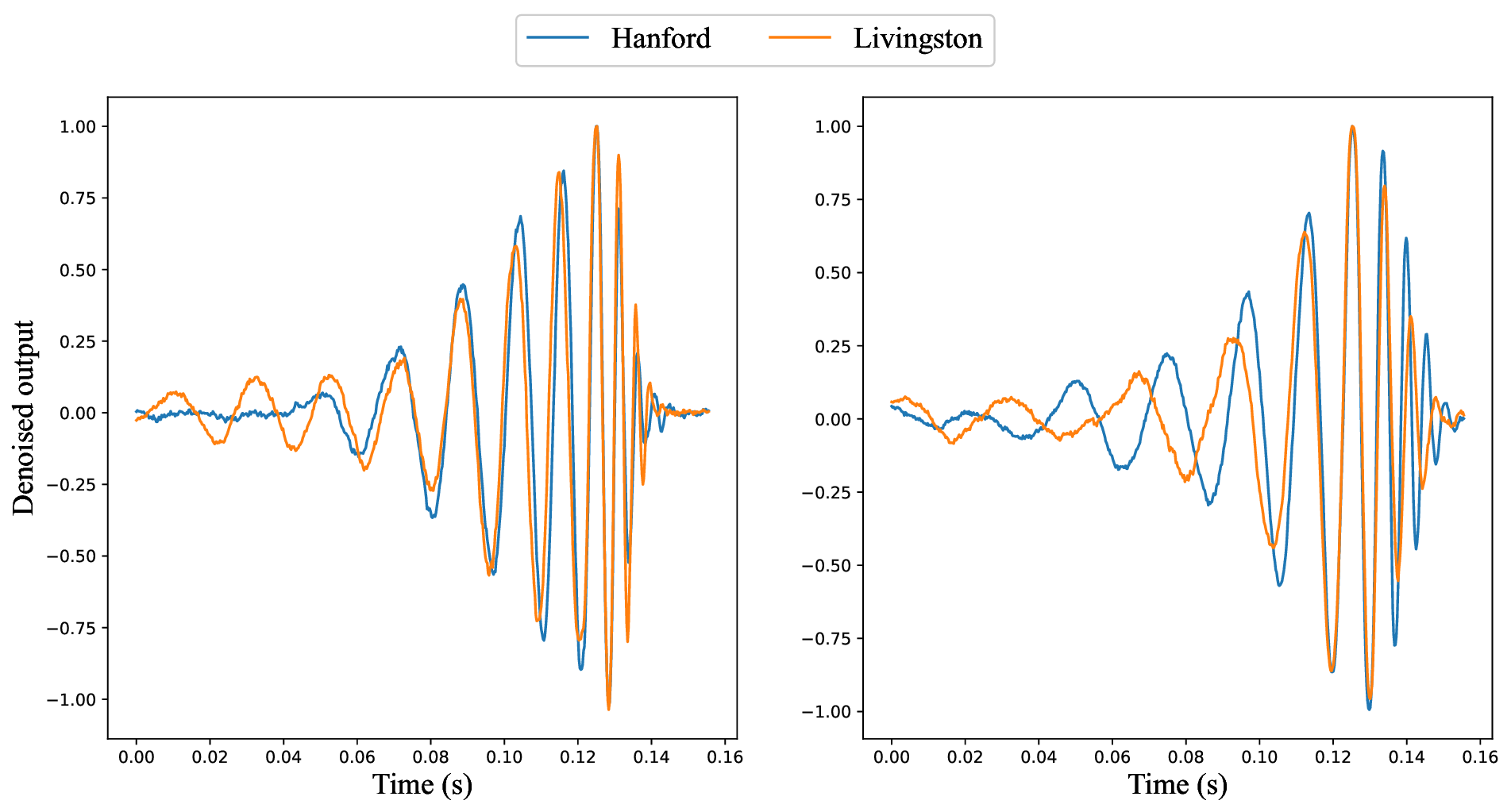}
		\caption{\label{shapeCompare} Comparation of the two false triggers’ denoised output of Hanford and Livingston. Left: The denoised outputs near GPS time 1187014846. Right: The denoised outputs near GPS time 1187612152.}
    \end{figure*}

	\subsection{\label{sec:level2} Results for GW search based on half month’s data}
	
	The previous experiments indicate the effectiveness of the proposed framework for GW search on test datasets and confident events in GWTC-1, GWTC-2.0, GWTC-2.1, and GWTC-3. Nevertheless, these previous findings may not adequately capture the framework's performance in long-time-duration detection scenarios. In this sub-section, we will delve into the framework's ability to detect strains spanning a half-month duration in August 2017.
	
	The data was used in the previous research \cite{28,31}. Because a section of the first half of the August 2017 strain was employed for training the astrophysical origin discrimination network, we only test the strain of the second half part, and the GPS time of the strains falls between 1186683883 and 1187733597. To prevent redundancy from earlier sections, our focus here is solely on examining the ``noise" background's response. Consequently, the confident events that occurred in August 2017 are not subjected to investigation within this sub-section.
	
	Table~\ref{table5} presents the experiment results of false triggers of the half-month strain. In the scenario where only the Hanford data is considered, Method I generates hundreds, and in some instances, thousands of false triggers. In contrast, Method III significantly reduces the number of false triggers compared to Method I. For instance, in the case $Mass1\in[5\ M_\odot,10\ M_\odot]$ and $Mass2\in[5\ M_\odot,10\ M_\odot]$, Method I yields 87 false triggers, whereas Method III produces only 2 false triggers. For the H+L case, Method II produces 15 false triggers (the same GPS time appearing in two or more parameter spaces we only count once). With the application of time difference testing, the number of false triggers decreases to 2 for Method II and 0 for Method III. 
	\begin{table*}[htbp]
		\centering
		\caption{The number of false triggers for the proposed framework detected by Hanford only (H) and detected by Hanford tested by Livingston (H+L) of Method I, II, and III. H, L, T, and S denote Hanford, Livingston, time difference testing, and network SNR threshold testing. The network SNR threshold is set to 8.}
		\begin{ruledtabular}
			\begin{tabular}{cccccccccccccc}
				\multirow{2}{*}{Mass1($\ M \odot$)}   &\multirow{2}{*}{Mass2($\ M \odot$)}     &\multicolumn{4}{c}{Method I}  & \multicolumn{4}{c}{Method II} &\multicolumn{4}{c}{Method III}  \\
				\cline{3-14}
				&				  &  H    & H+L  &H+L+T&H+L+T+S     &  H    & H+L & H+L+T &  H+L+T+S  & H   & H+L & H+L+T & H+L+T+S \\ 
				\hline  
				$[5,10]$	 &	$[5,10]$        &  87   & 1    &0    &0           &3      &0    &0      &0          &2    &0    &0      &0    \\   
				$[5,10]$     &	$[10,20]$       &  383  & 9    &0    &0           &135    &0    &0      &0          &29   &0    &0      &0    \\ 
				$[5,10]$	 &	$[20,40]$       &  1196 & 19   &0    &0           &284    &1    &0      &0          &44   &0    &0      &0        \\  
				$[5,10]$	 &	$[40,80]$       &  3265 & 32   &2    &2           &300    &0    &0      &0          &49   &0    &0      &0        \\  
				$[10,20]$	 &	$[10,20]$       &  717  & 5    &0    &0           &280    &0    &0      &0          &27   &0    &0      &0        \\  
				$[10,20]$	 &	$[20,40]$       &  2710 & 39   &1    &1           &646    &0    &0      &0          &34   &0    &0      &0        \\  
				$[10,20]$	 &	$[40,80]$       &  5259 & 55   &7    &7           &2290   &6    &1      &1          &92   &0    &0      &0        \\  
				$[20,40]$	 &	$[20,40]$       &  3823 & 41   &4    &4           &1197   &2    &0      &0          &45   &0    &0      &0        \\  
				$[20,40]$	 &	$[40,80]$       &  5185 & 60   &15   &11          &2238   &7    &0      &0          &129  &0    &0      &0        \\  
				$[40,80]$	 &	$[40,80]$       &  4375 & 135  &65   &3           &1061   &2    &1      &1          &246  &1    &0      &0        \\  
			\end{tabular}
		\end{ruledtabular}
		\label{table5}
	\end{table*}
	
	We believe that certain false triggers generated by our proposed framework warrant further investigation. We plot the information of the two false triggers generated by Method II which is shown in Fig.~\ref{strainCompare}. For easier comparison, the whitened strain is rescaled by 1/50. These two false triggers are characterized by GPS times approximately close to 1187014846 and 1187612152. Denoised results of the two false triggers exhibit chirp-like shapes. The SNRs calculated through matched filtering between the denoised results, and the whitened strain are all above 5.
	
	Displayed in Fig.~\ref{shapeCompare} is a comparison between the denoising outputs from Hanford and Livingston for the two false triggers.  The two signals denoised by the strain of Hanford and Livingston are shifted to align the time corresponding to the maximum value. Interestingly, the denoised shapes of Hanford and Livingston around GPS time 1187014846 appear remarkably similar. We calculated the overlap between the denoised output of the two interferometers near GPS time 1187014846 and yield 88.9\%. We believe that the strain near the GPS time can be further investigated in the future. This indicates the potential of our proposed framework for conducting more in-depth investigations into the archived data of GWTC-1, GWTC-2, and GWTC-3.
	
	\section{\label{sec:level1} Conclusion and Discussion}
	
	In this work, we introduce a novel framework for gravitational wave detection through matched filtering that operates without the necessity of a template bank. This framework employs an envelope extraction network alongside denoising networks and astrophysical origin discrimination networks. Notably, we've developed and trained 10 denoising networks and 10 astrophysical discrimination networks within this proposed framework. The test results of both the test set and real detector data demonstrate the effectiveness of the proposed framework.
	
	For the sake of simplicity in our analysis, we opted for a relatively compact neural network structure for denoising, consisting of just 11 layers. This denoising network can be executed and trained on a NVIDIA GeForce RTX 3060 Laptop GPU (6 GB). We demonstrate the effectiveness of neural network with relatively small scale for denoising. 
	
	We are of the opinion that more complex network architectures, such as deeper networks like WaveNet \cite{44}, wider networks like WaveFormer \cite{46}, and sequential modeled networks like CNN-LSTM \cite{45}, can be seamlessly integrated into our framework to enhance the overall gravitational wave search performance.
	
	The denoising output serves as an intermediary outcome within our approach. This intermediary result encompasses crucial details like arrival time delays, signal amplitudes, and phases. Such information can be harnessed to enhance various other facets of gravitational wave data processing, including tasks like localization and parameter estimation.
	
	Given the notably low phase recovery error exhibited by the end-to-end denoising model, our proposed method holds promise for effectively tackling the phase-related challenges in the search for Extreme Mass Ratio Inspirals (EMRIs) GW signals \cite{32} in the future. Additionally, this method exhibits potential utility in detecting GW signals stemming from binaries with quantifiable eccentricities.

	\section*{Acknowledgments}
	
	This research has made use of data and web tools obtained from the gravitational-wave Open Science Center, a service of LIGO Laboratory, the LIGO Scientific Collaboration, and the Virgo Collaboration.
	
	We thank Xinquan Chen for his help with the  data annotation. This work was supported in part by the National Key Research and Development Program of China Grant No. 2021YFC2203001, in part by the NSFC (No. 11920101003 and No. 12021003) and the Natural Science Foundation of Jiangxi (No. 20224BAB211012).
	
	Z. Cao was supported by CAS Project for Young Scientists in Basic Research YSBR-006.
	
	\nocite{*}
	\bibliography{paper}

\begin{thebibliography}{10}

\bibitem{1}
BP~Abbott, Richard Abbott, TDea Abbott, S~Abraham, F~Acernese, K~Ackley,
  C~Adams, RX~Adhikari, VB~Adya, Christoph Affeldt, et~al.
\newblock Gwtc-1: a gravitational-wave transient catalog of compact binary
  mergers observed by ligo and virgo during the first and second observing
  runs.
\newblock {\em Physical Review X}, 9(3):031040, 2019.

\bibitem{2}
R~Abbott, TD~Abbott, S~Abraham, F~Acernese, K~Ackley, A~Adams, C~Adams,
  RX~Adhikari, VB~Adya, Christoph Affeldt, et~al.
\newblock Gwtc-2: compact binary coalescences observed by ligo and virgo during
  the first half of the third observing run.
\newblock {\em Physical Review X}, 11(2):021053, 2021.

\bibitem{3}
R~Abbott, TD~Abbott, F~Acernese, K~Ackley, C~Adams, N~Adhikari, RX~Adhikari,
  VB~Adya, C~Affeldt, D~Agarwal, et~al.
\newblock Gwtc-3: compact binary coalescences observed by ligo and virgo during
  the second part of the third observing run.
\newblock {\em arXiv preprint arXiv:2111.03606}, 2021.

\bibitem{4}
R~Abbott, TD~Abbott, F~Acernese, K~Ackley, C~Adams, N~Adhikari, RX~Adhikari,
  VB~Adya, C~Affeldt, D~Agarwal, et~al.
\newblock Population of merging compact binaries inferred using gravitational
  waves through gwtc-3.
\newblock {\em Physical Review X}, 13(1):011048, 2023.

\bibitem{5}
Benjamin~P Abbott, R~Abbott, TD~Abbott, Fausto Acernese, K~Ackley, C~Adams,
  T~Adams, Paolo Addesso, Rana~X Adhikari, Vaishali~B Adya, et~al.
\newblock Tests of general relativity with gw170817.
\newblock {\em Physical review letters}, 123(1):011102, 2019.

\bibitem{6}
BP~Abbott, R~Abbott, TD~Abbott, F~Acernese, K~Ackley, C~Adams, T~Adams,
  P~Addesso, RX~Adhikari, VB~Adya, et~al.
\newblock A gravitational-wave standard siren measurement of the hubble
  constant.
\newblock {\em NATURE}, 551(7678):85, 2017.

\bibitem{7}
R~Abbott, TD~Abbott, S~Abraham, F~Acernese, K~Ackley, A~Adams, C~Adams,
  RX~Adhikari, VB~Adya, C~Affeldt, et~al.
\newblock Constraints on cosmic strings using data from the third advanced
  ligo--virgo observing run.
\newblock {\em Physical review letters}, 126(24):241102, 2021.

\bibitem{8}
R~Abbott, H~Abe, F~Acernese, K~Ackley, N~Adhikari, RX~Adhikari, VK~Adkins,
  VB~Adya, C~Affeldt, D~Agarwal, et~al.
\newblock All-sky search for continuous gravitational waves from isolated
  neutron stars using advanced ligo and advanced virgo o3 data.
\newblock {\em Physical Review D}, 106(10):102008, 2022.

\bibitem{9}
Alba Romero-Rodriguez, Mario Martinez, Oriol Pujolas, Mairi Sakellariadou, and
  Ville Vaskonen.
\newblock Search for a scalar induced stochastic gravitational wave background
  in the third ligo-virgo observing run.
\newblock {\em Physical Review Letters}, 128(5):051301, 2022.

\bibitem{10}
Quentin Baghi, Natalia Korsakova, Jacob Slutsky, Eleonora Castelli, Nikolaos
  Karnesis, and Jean-Baptiste Bayle.
\newblock Detection and characterization of instrumental transients in lisa
  pathfinder and their projection to lisa.
\newblock {\em Physical Review D}, 105(4):042002, 2022.

\bibitem{11}
Zheng-Cheng Liang, Yi-Ming Hu, Yun Jiang, Jun Cheng, Jian-dong Zhang, Jianwei
  Mei, et~al.
\newblock Science with the tianqin observatory: Preliminary results on
  stochastic gravitational-wave background.
\newblock {\em Physical Review D}, 105(2):022001, 2022.

\bibitem{12}
Ziren Luo, Yan Wang, Yueliang Wu, Wenrui Hu, and Gang Jin.
\newblock The taiji program: A concise overview.
\newblock {\em Progress of Theoretical and Experimental Physics},
  2021(5):05A108, 2021.

\bibitem{13}
Michele Maggiore, Chris Van~Den Broeck, Nicola Bartolo, Enis Belgacem, Daniele
  Bertacca, Marie~Anne Bizouard, Marica Branchesi, Sebastien Clesse, Stefano
  Foffa, Juan Garc{\'\i}a-Bellido, et~al.
\newblock Science case for the einstein telescope.
\newblock {\em Journal of Cosmology and Astroparticle Physics},
  2020(03):050--050, 2020.

\bibitem{14}
Evan~D Hall.
\newblock Cosmic explorer: A next-generation ground-based gravitational-wave
  observatory.
\newblock {\em Galaxies}, 10(4):90, 2022.

\bibitem{15}
Christopher~Michael Biwer, Collin~D Capano, Soumi De, Miriam Cabero, Duncan~A
  Brown, Alexander~H Nitz, and Vivien Raymond.
\newblock Pycbc inference: A python-based parameter estimation toolkit for
  compact binary coalescence signals.
\newblock {\em Publications of the Astronomical Society of the Pacific},
  131(996):024503, 2019.

\bibitem{16}
Kipp Cannon, Sarah Caudill, Chiwai Chan, Bryce Cousins, Jolien~DE Creighton,
  Becca Ewing, Heather Fong, Patrick Godwin, Chad Hanna, Shaun Hooper, et~al.
\newblock Gstlal: A software framework for gravitational wave discovery.
\newblock {\em SoftwareX}, 14:100680, 2021.

\bibitem{17}
Florian Aubin, Francesco Brighenti, Roberto Chierici, Dimitri Estevez, Giuseppe
  Greco, Gianluca~Maria Guidi, Vincent Juste, Fr{\'e}d{\'e}rique Marion, Benoit
  Mours, Elisa Nitoglia, et~al.
\newblock The mbta pipeline for detecting compact binary coalescences in the
  third ligo--virgo observing run.
\newblock {\em Classical and Quantum Gravity}, 38(9):095004, 2021.

\bibitem{18}
Qi~Chu, Manoj Kovalam, Linqing Wen, Teresa Slaven-Blair, Joel Bosveld, Yanbei
  Chen, Patrick Clearwater, Alex Codoreanu, Zhihui Du, Xiangyu Guo, et~al.
\newblock Spiir online coherent pipeline to search for gravitational waves from
  compact binary coalescences.
\newblock {\em Physical Review D}, 105(2):024023, 2022.

\bibitem{19}
Pablo~J Barneo, Alejandro Torres-Forn{\'e}, Jos{\'e}~A Font, Marco Drago, Jordi
  Portell, and Antonio Marquina.
\newblock Implementation of the regularized rudin-osher-fatemi denoising method
  in the coherent wave burst pipeline for gravitational-wave data analysis.
\newblock {\em Physical Review D}, 106(2):022002, 2022.

\bibitem{20}
Daniel George and Eliu~Antonio Huerta.
\newblock Deep learning for real-time gravitational wave detection and
  parameter estimation: Results with advanced ligo data.
\newblock {\em Physics Letters B}, 778:64--70, 2018.

\bibitem{21}
Hunter Gabbard, Michael Williams, Fergus Hayes, and Chris Messenger.
\newblock Matching matched filtering with deep networks for gravitational-wave
  astronomy.
\newblock {\em Physical review letters}, 120(14):141103, 2018.

\bibitem{22}
Man~Leong Chan, Ik~Siong Heng, and Chris Messenger.
\newblock Detection and classification of supernova gravitational wave signals:
  A deep learning approach.
\newblock {\em Physical Review D}, 102(4):043022, 2020.

\bibitem{23}
Gr{\'e}gory Baltus, Justin Janquart, Melissa Lopez, Amit Reza, Sarah Caudill,
  and Jean-Ren{\'e} Cudell.
\newblock Convolutional neural networks for the detection of the early inspiral
  of a gravitational-wave signal.
\newblock {\em Physical Review D}, 103(10):102003, 2021.

\bibitem{24}
Banafsheh Beheshtipour and Maria~Alessandra Papa.
\newblock Deep learning for clustering of continuous gravitational wave
  candidates. ii. identification of low-snr candidates.
\newblock {\em Physical Review D}, 103(6):064027, 2021.

\bibitem{25}
Rich Ormiston, Tri Nguyen, Michael Coughlin, Rana~X Adhikari, and Erik
  Katsavounidis.
\newblock Noise reduction in gravitational-wave data via deep learning.
\newblock {\em Physical Review Research}, 2(3):033066, 2020.

\bibitem{26}
Chayan Chatterjee, Linqing Wen, Kevin Vinsen, Manoj Kovalam, and Amitava Datta.
\newblock Using deep learning to localize gravitational wave sources.
\newblock {\em Physical Review D}, 100(10):103025, 2019.

\bibitem{27}
He~Wang, Shichao Wu, Zhoujian Cao, Xiaolin Liu, and Jian-Yang Zhu.
\newblock Gravitational-wave signal recognition of ligo data by deep learning.
\newblock {\em Physical Review D}, 101(10):104003, 2020.

\bibitem{28}
CunLiang Ma, Wei Wang, He~Wang, and Zhoujian Cao.
\newblock Ensemble of deep convolutional neural networks for real-time
  gravitational wave signal recognition.
\newblock {\em Physical Review D}, 105(8):083013, 2022.

\bibitem{29}
Marlin~B Sch{\"a}fer and Alexander~H Nitz.
\newblock From one to many: A deep learning coincident gravitational-wave
  search.
\newblock {\em Physical Review D}, 105(4):043003, 2022.

\bibitem{30}
Marlin~B Sch{\"a}fer, Ond{\v{r}}ej Zelenka, Alexander~H Nitz, Frank Ohme, and
  Bernd Br{\"u}gmann.
\newblock Training strategies for deep learning gravitational-wave searches.
\newblock {\em Physical Review D}, 105(4):043002, 2022.

\bibitem{31}
Cunliang Ma, Wei Wang, He~Wang, and Zhoujian Cao.
\newblock Artificial intelligence model for gravitational wave search based on
  the waveform envelope.
\newblock {\em Physical Review D}, 107(6):063029, 2023.

\bibitem{32}
Xue-Ting Zhang, Chris Messenger, Natalia Korsakova, Man~Leong Chan, Yi-Ming Hu,
  and Jian-dong Zhang.
\newblock Detecting gravitational waves from extreme mass ratio inspirals using
  convolutional neural networks.
\newblock {\em Physical Review D}, 105(12):123027, 2022.

\bibitem{33}
Christoph Dreissigacker, Rahul Sharma, Chris Messenger, Ruining Zhao, and
  Reinhard Prix.
\newblock Deep-learning continuous gravitational waves.
\newblock {\em Physical Review D}, 100(4):044009, 2019.

\bibitem{34}
M~L{\'o}pez, I~Di~Palma, M~Drago, P~Cerd{\'a}-Dur{\'a}n, and F~Ricci.
\newblock Deep learning for core-collapse supernova detection.
\newblock {\em Physical Review D}, 103(6):063011, 2021.

\bibitem{35}
Seiya Sasaoka, Yilun Hou, Kentaro Somiya, and Hirotaka Takahashi.
\newblock Localization of gravitational waves using machine learning.
\newblock {\em Physical Review D}, 105(10):103030, 2022.

\bibitem{36}
Chayan Chatterjee, Linqing Wen, Kevin Vinsen, Manoj Kovalam, and Amitava Datta.
\newblock Using deep learning to localize gravitational wave sources.
\newblock {\em Physical Review D}, 100(10):103025, 2019.

\bibitem{37}
Maximilian Dax, Stephen~R Green, Jonathan Gair, Jakob~H Macke, Alessandra
  Buonanno, and Bernhard Sch{\"o}lkopf.
\newblock Real-time gravitational wave science with neural posterior
  estimation.
\newblock {\em Physical review letters}, 127(24):241103, 2021.

\bibitem{38}
He~Wang, Zhoujian Cao, Yue Zhou, Zong-Kuan Guo, and Zhixiang Ren.
\newblock Sampling with prior knowledge for high-dimensional gravitational wave
  data analysis.
\newblock {\em Big Data Mining and Analytics}, 5(1):53--63, 2021.

\bibitem{39}
Jurriaan Langendorff, Alex Kolmus, Justin Janquart, and Chris Van Den~Broeck.
\newblock Normalizing flows as an avenue to studying overlapping gravitational
  wave signals.
\newblock {\em Physical Review Letters}, 130(17):171402, 2023.

\bibitem{40}
Hunter Gabbard, Chris Messenger, Ik~Siong Heng, Francesco Tonolini, and
  Roderick Murray-Smith.
\newblock Bayesian parameter estimation using conditional variational
  autoencoders for gravitational-wave astronomy.
\newblock {\em Nature Physics}, 18(1):112--117, 2022.

\bibitem{41}
Huilan Luo, Pei Wang, Hongkun Chen, and Min Xu.
\newblock Object detection method based on shallow feature fusion and semantic
  information enhancement.
\newblock {\em IEEE Sensors Journal}, 21(19):21839--21851, 2021.

\bibitem{42}
Shuxin Yang, Suxin Tong, Guixiang Zhu, Jie Cao, Youquan Wang, Zhengfa Xue,
  Hongliang Sun, and Yu~Wen.
\newblock Mve-flk: A multi-task legal judgment prediction via multi-view
  encoder fusing legal keywords.
\newblock {\em Knowledge-Based Systems}, 239:107960, 2022.

\bibitem{43}
Timothy~D Gebhard, Niki Kilbertus, Ian Harry, and Bernhard Sch{\"o}lkopf.
\newblock Convolutional neural networks: A magic bullet for gravitational-wave
  detection?
\newblock {\em Physical Review D}, 100(6):063015, 2019.

\bibitem{44}
Wei Wei and EA~Huerta.
\newblock Gravitational wave denoising of binary black hole mergers with deep
  learning.
\newblock {\em Physics Letters B}, 800:135081, 2020.

\bibitem{45}
Chayan Chatterjee, Linqing Wen, Foivos Diakogiannis, and Kevin Vinsen.
\newblock Extraction of binary black hole gravitational wave signals from
  detector data using deep learning.
\newblock {\em Physical Review D}, 104(6):064046, 2021.

\bibitem{46}
Zhixiang Ren, He~Wang, Yue Zhou, Zong-Kuan Guo, and Zhoujian Cao.
\newblock Intelligent noise suppression for gravitational wave observational
  data.
\newblock {\em arXiv preprint arXiv:2212.14283}, 2022.

\bibitem{47}
Alexander~H Nitz, Thomas Dent, Tito Dal~Canton, Stephen Fairhurst, and Duncan~A
  Brown.
\newblock Detecting binary compact-object mergers with gravitational waves:
  Understanding and improving the sensitivity of the pycbc search.
\newblock {\em The Astrophysical Journal}, 849(2):118, 2017.

\bibitem{48}
Samantha~A Usman, Alexander~H Nitz, Ian~W Harry, Christopher~M Biwer, Duncan~A
  Brown, Miriam Cabero, Collin~D Capano, Tito Dal~Canton, Thomas Dent, Stephen
  Fairhurst, et~al.
\newblock The pycbc search for gravitational waves from compact binary
  coalescence.
\newblock {\em Classical and Quantum Gravity}, 33(21):215004, 2016.

\bibitem{49}
Hua-Mei Luo, Wenbin Lin, Zu-Cheng Chen, and Qing-Guo Huang.
\newblock Extraction of gravitational wave signals with optimized convolutional
  neural network.
\newblock {\em Frontiers of Physics}, 15:1--6, 2020.

\bibitem{50}
Heming Xia, Lijing Shao, Junjie Zhao, and Zhoujian Cao.
\newblock Improved deep learning techniques in gravitational-wave data
  analysis.
\newblock {\em Physical Review D}, 103(2):024040, 2021.

\end{thebibliography}
	\end{document}